\documentclass[table,xcdraw,12pt]{aastex63}
\usepackage{xcolor}

\usepackage{tabularx}
\usepackage{colortbl}
\usepackage{array}
\usepackage{wrapfig}
\usepackage{mwe}
\usepackage{comment}
\usepackage{nicematrix}
\usepackage{makecell}
\usepackage{float}
\usepackage{multirow}
\usepackage{url} 
\def\UrlBreaks{\do\/\do-} 
\arrayrulecolor[HTML]{560319} 
\newlength{\thickarrayrulewidth}
\renewcommand*{\arraystretch}{2} 
\shorttitle{Predicting Solar Proton Events with GOES}
\shortauthors{Ali et al.}
\graphicspath{{./figures/}}

\begin{document}

\title{\centering Predicting Solar Proton Events of Solar Cycles 22 – 24 using GOES \\ Proton \& soft X-ray flux features }
 
\correspondingauthor{Aatiya Ali}
\email{aali87@student.gsu.edu}

\author[0000-0003-3196-3822]{Aatiya Ali}
\affiliation{Physics \& Astronomy Department, Georgia State University, Atlanta, GA 30303, USA}

\author[0000-0002-4001-1295]{Viacheslav Sadykov}
\affiliation{Physics \& Astronomy Department, Georgia State University, Atlanta, GA 30303, USA}  

\author[0000-0003-0364-4883]{Alexander Kosovichev} 
\affiliation{Physics Department, New Jersey Institute of Technology, Newark, NJ 07102, USA}
\affiliation{NASA Ames Research Center, Moffett Field, CA 94035, USA}

\author[0000-0003-4144-2270]{Irina N. Kitiashvili}
\affiliation{NASA Ames Research Center, Moffett Field, CA 94035, USA}

\author{Vincent Oria}
\affiliation{Computer Science Department, New Jersey Institute of Technology, Newark, NJ 07102, USA}

\author[0000-0003-2846-2453]{Gelu M. Nita}
\affiliation{Physics Department, New Jersey Institute of Technology, Newark, NJ 07102, USA}

\author[0000-0002-2858-9625]{Egor Illarionov}
\affiliation{Department of Mechanics and Mathematics, Moscow State University, Moscow, 119991, Russia}
\affiliation{Moscow Center of Fundamental and Applied Mathematics, Moscow, 119234, Russia}

\author{Patrick M. O’Keefe}
\affiliation{Computer Science Department, New Jersey Institute of Technology, Newark, NJ 07102, USA}

\author{Fraila Francis}
\affiliation{Computer Science Department, New Jersey Institute of Technology, Newark, NJ 07102, USA}

\author{Chun-Jie Chong}
\affiliation{Computer Science Department, New Jersey Institute of Technology, Newark, NJ 07102, USA}

\author{Paul Kosovich}
\affiliation{Physics Department, New Jersey Institute of Technology, Newark, NJ 07102, USA}

\author{Russell D. Marroquin}
\affiliation{Department of Physics, University of California San Diego, La Jolla, CA 92093, USA}

\begin{abstract}
    Solar Energetic Particle (SEP) events and their major subclass, Solar Proton Events (SPEs), can \textnormal{have unfavorable consequences on} numerous aspects of life and technology, making them one of the most harmful effects of solar activity. \textnormal{Garnering} knowledge preceding such events by studying \textnormal{operational data flows is essential for their forecasting.} \textnormal{Considering only Solar Cycle (SC) 24 in our previous study \citep{slava2021}, we found that it may be sufficient to utilize only proton and soft X-ray (SXR) parameters for SPE forecasts.} \textnormal{Here, we} report a catalog recording $\geq$10 MeV $\geq$10 particle flux unit SPEs \textnormal{with their properties, spanning SCs 22--24,} using NOAA's Geostationary Operational Environmental Satellite flux data. We report an additional catalog of daily proton and SXR flux statistics \textnormal{for this period, employing it to} test the application of machine learning \textnormal{(ML) on} the prediction of SPEs \textnormal{using a Support Vector Machine (SVM) and eXtreme Gradient Boosting (XGBoost). We explore the effects of training models with data from one \textit{and} two SCs, evaluating how transferable a model can be across different time periods. XGBoost proved to be more accurate than SVMs for almost every test considered, while outperforming operational SWPC NOAA predictions and a persistence forecast. Interestingly, training done with SC 24 produces weaker TSS and HSS$_{2}$, even when paired with SC 22 or SC 23, indicating transferability issues. This work contributes towards validating forecasts using long-spanning data-- an understudied area in SEP research that should be considered to verify the cross-cycle robustness of ML-driven forecasts.}

\end{abstract}

;

\section{Introduction} \label{sec:intro}
Solar Energetic Particle (SEP) events are enhanced fluxes of high-energy particles ejected by the Sun. The occurrence rates of such events are greatest \textnormal{closer to} the maxima of $\sim$11-year Solar Cycles (SCs). \textnormal{These events} encompass a wide range of energies from \textnormal{KeVs up to multiple GeVs} \citep{anasta2019}, ejected into the heliosphere. Solar Proton Events (SPEs), a subclass of SEPs, are \textnormal{characterized as protons with energies $\geq$10 MeV exceeding a threshold of $\geq$10 particle flux units (pfus).} Energetic protons can harm satellites, \textnormal{navigation-communication systems, technological grids,} and other equipment. High-energy charged particles in the magnetosphere can manipulate \textnormal{the output} signals of electronic devices. \textnormal{This causes spacecrafts' calibration systems to \textnormal{malfunction}, and when} these charged particles strike a critical device, the instrument may fail entirely. \textnormal{Other examples of solar transient activity and its consequences include high-energy electrons that further complicate operations as they \textnormal{can} penetrate shielding aboard satellites and spacecraft. They quickly pile up, and eventually, discharge the accumulated energy as a ``lightning strike'' \citep{lsp}. Radio-wave-dependent} communication systems are also vulnerable to such events. \textnormal{Extreme Ultraviolet and X-ray radiation from the Sun can ionize the Earth's ionosphere, elevating electron density in the medium radio waves travel through. This delays transmission time from satellites to ground-based global positioning systems, ultimately causing the misalignment of positions by a few meters.} While seemingly insignificant, this poses issues for aviation, robotics, military, transportation, and other industries' operations \citep{damaging_satellites}. \textnormal{SEPs and cosmic rays are similarly capable of ionizing and altering the ionosphere \citep{iono}.}

\textnormal{The limited understanding of solar processes leading to SPE generation motivates research in Heliophysics and astrobiology to advance. By enhancing the \textnormal{radiation levels} in interplanetary space \citep{langford_2022}, SPEs may be responsible for the atrophy of astronaut health. \citet{bio1} expand on this, highlighting the risks \textnormal{of} astronauts developing cancer, experiencing central nervous system decrements, and even exhibiting degenerative tissue effects. More recently, such concerns have caught the attention of administrations working with commercial airlines and space tourism. \citet{planes2005} conclude that combined effects of magnetic field disturbances and solar particle fluence due to solar storms can be responsible for up to $\sim$70\% variation in radiation exposure at typical flight altitudes. \citet{tourism} also notes higher exposure to cosmic radiation as the main safety concern during space tourism between lunar and orbital travel. This would only be enhanced during the propagation of SEPs. Further, \citet{shielding} conclude that some composite materials \citep[i.e. carbon fiber-reinforced plastic \& silicon-carbon plastic,][]{lsp} show promise in shielding spacecraft from enhanced radiation. Still, none have been confirmed to withstand the expected \textnormal{radiative} variations during SPEs. Given the \textnormal{many hazards presented by these events, it is critical to develop reliable forecasts to provide sufficient} time for astronauts and equipment to be safely relocated. External SEP detectors coupled with an appropriate predictive algorithm may be mounted onto spacecraft to act as warning and defense systems in this effort. Our overarching goal is to investigate the capabilities of different machine learning (ML) models with respect to the prediction of SPEs. We employ our models with only proton and soft X-ray (SXR) flux data, and assess performance variations when considering timescales longer than a single SC during the training phase.}

\subsection{The problem of Solar Proton Event prediction} \label{sec:problem}
\begin{figure}[t!]
    \centering
    \includegraphics[width=.75\linewidth]{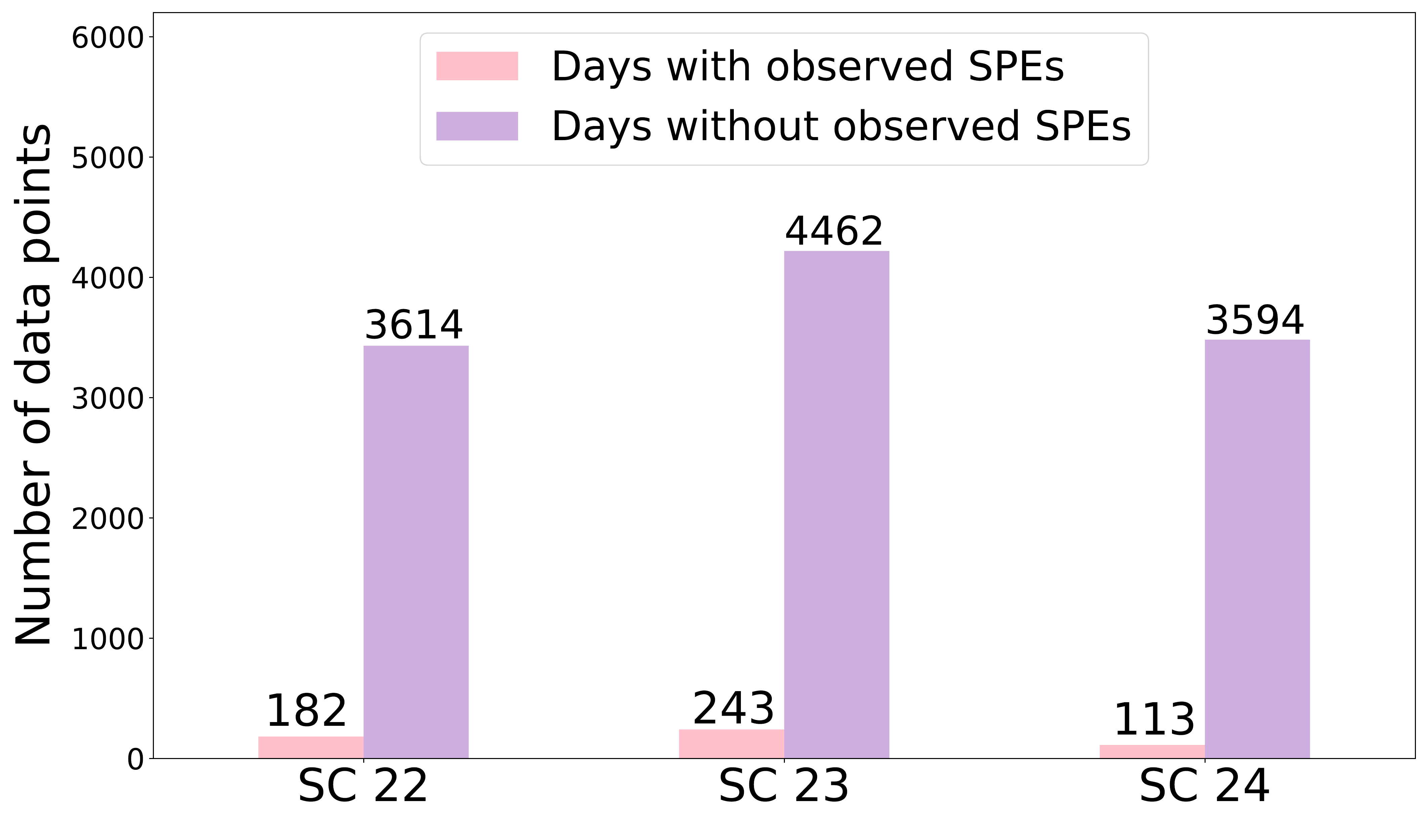}
    \caption{The drastic imbalance \textnormal{in the number of days} where SPEs are observed per SC, compared to days they are not observed. \textit{Note, SC 22 is dated from \textnormal{September 1986 to August 1996, SC 23 from August 1996 to December 2008, and SC 24 from December 2008 to December 2019.}}}
    \label{fig:imbalance}
\end{figure}

Approaches to building predictive models for SPEs become arduous given their rare nature and reliance on indirect measures of volatile SC activity. If using measures like forecasting accuracy to quantify the success of a predictive model, the rarity of SPEs permits an algorithm to be able to miss \textnormal{most, if not all, incoming SPEs while sustaining high accuracy in prediction scores (as we see in Section \ref{sec:eval})}. The need for balanced data \textnormal{sets and assessment metrics to reflect accurate scores are} emphasized by \citet{martens2017}. Specifically, the class imbalance in our data set shows 11946 negative cases (days with no SPEs), in vast contrast to only 538 positive cases (days with SPEs). This imbalance overshadows positive cases by the prevalence of quiet \textnormal{(days with no observed SPEs) periods, and is} the overarching problem encountered in this work, and the disparity is visualized in Fig. \ref{fig:imbalance}. \textnormal{Variance in the Sun's global magnetic activity and consequential changing event frequency brings forward the question of how transferable an algorithm built using previous SC data may be when considering future SCs of unknown activity levels.}

\subsection{Current \& previous \textnormal{prediction efforts} and limitations} \label{sec:lit_rev}
In recent years, there has been a plethora of research projects contributing to the effort of predicting SPEs in an attempt to mitigate their detrimental effects. In our previous study \citep{slava2021} we use SXR wavelength ranges (long (0.1 - 0.8 nm) and short (0.05 - 0.4 nm)), along with $\geq$10 MeV proton flux data observed by the Geostationary Operational Environmental Satellite (GOES) series. From the various products obtained by GOES, we retrieve and use SXR flux data with a 1-minute cadence and $\geq$10 MeV proton flux data with a 5-minute cadence. The data \textnormal{have} been made publicly available by the National Oceanic \& Atmospheric Association (NOAA) National Center for Environmental Information (NCEI \footnote{\url{www.ncei.noaa.gov/data/goes-space-environment-monitor/access/avg/}}). The \textnormal{promising results} of utilizing derivatives of these data products alone \textnormal{motivates} us to continue exploring these parameters in depth in this work. \citet{slava2021} also discuss the lack of performance loss when \textnormal{excluding characteristics} of Active Regions (ARs) and type II, III, \textnormal{and IV} radio bursts when generating predictive algorithms, although acknowledging the \textnormal{brevity} of the considered data set \textnormal{(SC 24 alone). Predictive scores resulting from using proton flux alone were compared to those with the addition of SXR data; which} proved to enhance prediction accuracy. \textnormal{Therefore, in this work we study operational proton and SXR flux features in detail, exploring the potential to develop a reliable SPE predictive model using these features.}

However, in addition to these parameters, different approaches to SPE prediction include physics-based and empirical models that take into account parameters of solar magnetograms, optical imaging, extreme ultraviolet (EUV) imaging, CMEs from single or multiple vantage points, in situ energetic proton and electron observations, particle acceleration and transport, \textnormal{and measurements} of solar wind density, temperature, and magnetic fields \citep{whitman2022}. \citet{whitman2022} also elaborate on current model validation diagnostics. \textnormal{Most physics-based models (i.e. computing SEP acceleration and transport from first principles) are computationally expensive, limiting their integration into current workflows to make ``real-time'' predictions. Statistics-based or ML-driven models can capture empirical, yet non-linear dependencies between observational data and SPE processes given sufficient data presented for training these models. We supply short-term (data from a single SC) \textit{and} long-term (data from two SCs) fluxes during our models' learning phases, exploring proton, short, and long SXR variations leading to SPEs. We use data acquired using only NOAA's GOES series for consistency.}

Studies comparing results between ML-driven models with those used as daily operational forecasts by NOAA's Space Weather Prediction Center (SWPC) for SPEs may \textnormal{identify observed} parameters preceding SPEs that could improve such forecasts \citep{slava2021}. To achieve this, we aggregate a catalog of statistical \textnormal{proton, short, and long SXR flux} parameters (discussed in Section \textnormal{\ref{sec:cat2}}), which are minimized by meticulously selecting features \textnormal{most} relevant to predicting \textnormal{these SEP events during unpredictable levels of solar activity.} In parallel, fundamental parameters required by the model may be poorly characterized without a complete understanding of the underlying \textnormal{solar mechanisms} associated with SEP ejection and acceleration. \textnormal{SEP modeling has thus been motivated to explore the physical processes related to SEPs and operational} forecasting needs. These current complex models show promise \textnormal{in} modeling \textnormal{time-dependent distributions of SEP events.} ML approaches are still being investigated to yield a new class of SEP models to produce fast, reliable forecasts \citep{whitman2022}. Our work here \textnormal{presents a different} approach to the problem.  

\subsection{Scope of \textnormal{our work}} \label{sec:scope}

\textnormal{The \textnormal{presented} work assesses ML-driven models for the prediction of SPEs using data from three previous SCs (22 -- 24). From this, we build an understanding of the cross-cycle transferability of these pre-established models. Starting with detecting SPEs from these SCs using GOES series data, event parameters are stored in the first catalog built in this effort. Using the continuous flux records we have of these SCs, we form an additional catalog of daily statistical features of SXR and $\geq$10 MeV proton fluxes, supplying the input data sets for our learning models.}

\textnormal{Our first model, Support vector machines (SVMs),} have become standard classifiers in space weather prediction studies. Recent works by \citet{bobra}, \citet{azim2021}, \citet{kasapis}, \citet{svm-more}, \citet{bobra2016}, and others, employ these classifiers and acknowledge them being advantageous, robust, and fast when making predictions. \citet{bobra} also show that SVMs are successful classifiers when applied to large data sets. Although developed more recently than SVMs, eXTreme gradient boosting (XGBoost; our second model) has already gained recognition in capturing complex patterns in the data. Similar to SVMs, it is well-suited for analyzing large data sets and is currently used in many areas of research (finance, healthcare, environmental sciences, etc.). \citet{sumanth23}, \citet{xgb-use}, \citet{xgb-use2}, and \citet{xgb-use3} highlight the success of using XGBoost to analyze SEP events, predict flares, and explore other space weather events. Within these works, XGBoost often outperforms more complex models (random forest, logistic regression, etc.) \textnormal{when regarding} different model performance metrics.

\textnormal{We begin our work here by constructing SPE flux data catalogs, one of which is used as the input for our ML models (discussed in Sections \ref{sec:cat1} \& \ref{sec:cat2}). To combat the imbalance in the positive and negative classes, we use model-inherent class-weight balancing techniques, as well as \textnormal{data generation using standard (duplication of positive cases), and synthetic oversampling}. We reduce the number of features considered for forecasting using Gini Importance, Fisher-scoring, and XGBoost, exploring which supplied statistical features are most important when working towards making predictions. Finally, we define data from each SC as different training/testing data sets (discussed in Section \ref{sec:sc_sep}) to use with SVM and XGBoost algorithms. Predictive capabilities are then measured using True Skill Statistic (TSS), Heidke Skill Scores \citep[HSS$_{2}$, developed by][]{mason2010}, and recall \textnormal{metrics}. We also apply k-fold cross-validation (CV) to optimize our models with respect to TSS. We also consider the effects of different training timescales when producing forecasts. Analyzing different parameters associated with SVM and XGBoost models, we strive to generalize our algorithms while retaining the highest scores achievable from our data set. Modifications applied to each model are further discussed in Section \ref{sec:apply}.}

\section{GOES Data Preparation \textnormal{\&} Products} \label{sec:methods}
\subsection{Querying \textnormal{flux data}} \label{sec:getdata}
Our prediction \textnormal{algorithms are} built using $\geq$10 MeV proton and SXR flux data queried from NOAA's National Center for Environmental Information (NCEI) \textnormal{through} 1986 -- 2019 to encompass SCs 22 -- 24. The data are obtained by different GOES launched by NOAA as a series from GOES-05 to GOES-15 during the period of interest. \textnormal{The main objective of GOES is to aid forecasting operations by supplying real-time access to X-ray and proton flux measurements \citep{aminalragia2021} from the geostationary orbit (an altitude of $\sim$36,000 km above Earth's equator)}; making this \textnormal{satellite} series a clear \textnormal{choice for several SC-long} data collection for our work. In 1974, NOAA compiled a primary and secondary scheme\footnote{ \url{www.ngdc.noaa.gov/stp/satellite/goes/doc/GOES\_XRS\_readme.pdf}} identifying GOES data to utilize during instances when, at each time, multiple satellites in the GOES series provided real-time data, assigning one to be the ``primary'' and others as ``secondary.'' \textnormal{\citet{sumanth} also do this in an effort to explore integral proton flux intensity profiles for space weather predictions. Influenced by these routines, we manually choose a ``primary'' instrument for every month across SCs 22 -- 24. By our definition, a primary instrument reflects a higher peak proton flux count when two or more satellites or detectors are simultaneously capturing data. Other works prioritize GOES data differently, e.g. \citet{aminalragia2021} retain measurements from the highest-numbered GOES satellite instead.}

\begin{figure}
    \centering
    \includegraphics[width=1\linewidth]{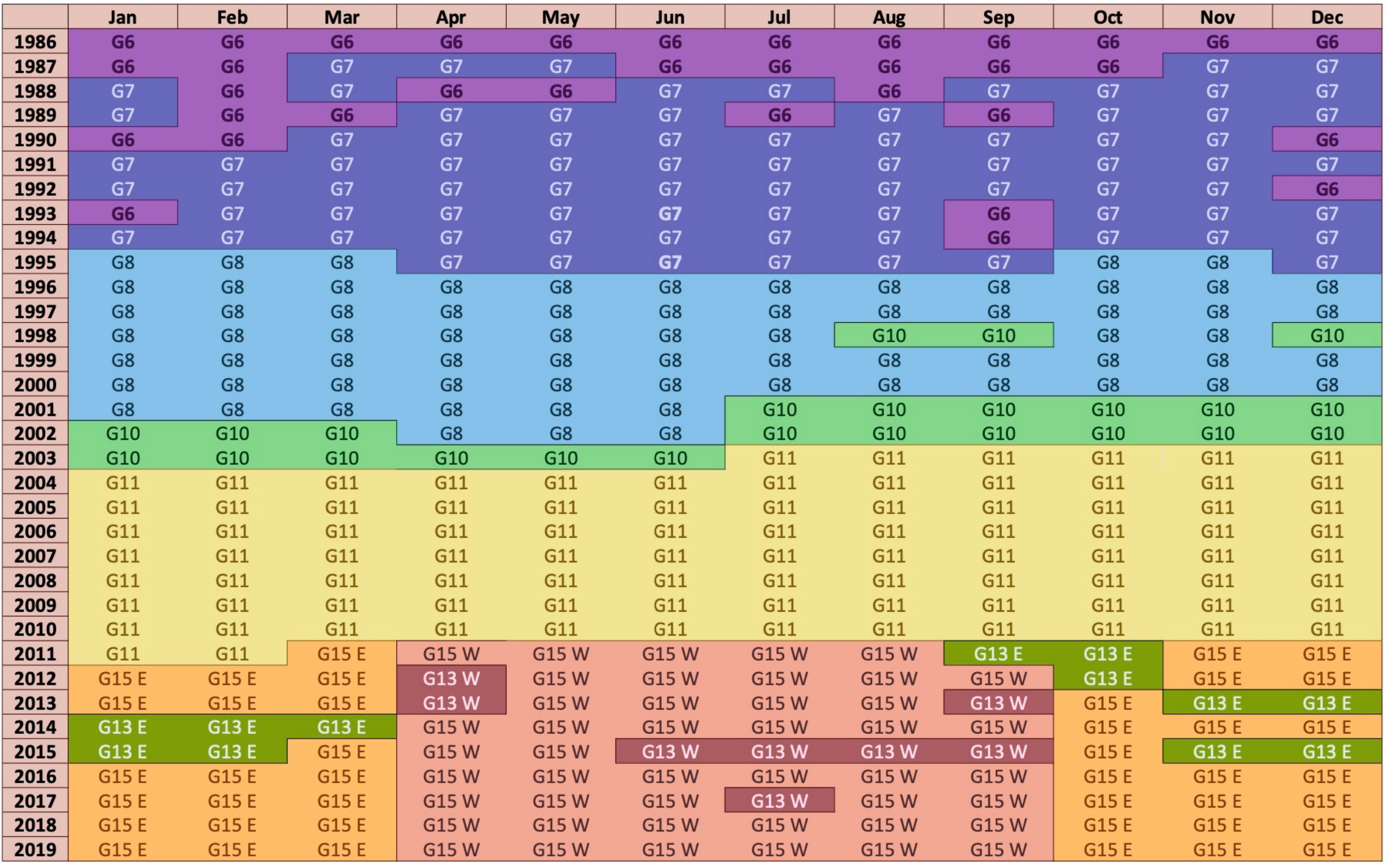}
    \caption{Timeline of \textnormal{``primary''} GOES instruments and detectors (when applicable) used to streamline SC 22 -- 24 flux data. Here, \textit{G} represents \textnormal{``GOES,'' succeeding} numbers represent the satellite number of the GOES series, and \textit{E} or \textit{W} (when applicable) indicates which of the East or West satellite detectors had detected the higher proton flux.}
    \label{fig:primaries}
\end{figure}

GOES \textnormal{underwent} a change from a mounted Energetic Particle Sensor (EPS) to an Energetic Proton, \textnormal{Electron, and Alpha Detector (EPEAD) in 2011 starting with GOES-13. Equipped with a single detector, the EPS was capable of distinguishing between SEPs and galactic cosmic rays, measuring fluxes from different energy channels. Upgrading this instrument,} EPEAD collects data from two detectors, one surveying East, at 75° W and the other West, at 135° W (with respect to the prime meridian); capturing slightly different populations of protons. \citet{rodriguez2014} further discuss this, concluding that differences in proton flux measurements between the detectors are due to the effect of magnetic field variation with geomagnetic longitude. This highlights that simply taking the average of these detectors’ fluxes will not accurately \textnormal{represent SEP propagation from the Sun to the magnetosphere.} Therefore, when EPEAD satellite data are introduced into our data set, both primary instruments and primary detectors (labeled as East or West) are selected and recorded for use. The primary instrument selection used for \textnormal{SCs 22 -- 24} is shown in \textnormal{Fig.} \ref{fig:primaries}, and a visualization comparing proton populations captured by the primary versus secondary detectors is presented in Fig. \ref{fig:prim-sec-comp}, reflecting the minimal- but still distinct differences in proton \textnormal{fluxes recorded by each detector.}
 
\begin{figure}[b!]
    \centering
    \includegraphics[width=.85\linewidth]{prim-sec-comp.pdf}
    \caption{\textnormal{Example of different proton populations detected by East and West detectors of GOES-13 \textnormal{for July} 2017 (shown are daily median fluxes). Given the consistently higher fluxes registered by the West detector compared to the East during the same observation period, we record it as the primary instrument/detector for the month.}}
    \label{fig:prim-sec-comp} 
\end{figure}

It is also important to note changes in GOES directionality after launch for those interested in recording instruments/detectors with no \textnormal{data gaps.} \textnormal{Specified} by \citet{rodriguez2012-flip} and the National Geophysical Data Center, GOES 13, \textnormal{14,} \& 15 all undergo a yaw flip where the satellite rotates about \textnormal{its} axis pointed toward the center of the \textnormal{Earth, flipping detector orientations. During these flips, EPEAD telemetry channels labeled ``East'' are actually looking westward, and those labeled ``West'' are looking eastward}. The inversions in directionality over time are as follows:

\begin{itemize} 
  \item GOES-13 (2006 -- 2018): only upright during its operational period between May 2010 and September 2012.  
  \item GOES-14 (2009 -- 2020): upright from its launch date in 2009 and inverted during \textnormal{an SPE} in early September 2012. The satellite has not corrected itself since.\footnote{Powered off in 2020, this satellite can be called back into service if needed \citep{noaa_2019}.}
  \item GOES-15 (2010 -- 2018): experiences a flip twice a year at every equinox. This maneuver usually lasts under an hour, during which data are not recorded. 
\end{itemize}

These inversions are considered for each satellite by their specific guidelines, and orientation labels are corrected for in the catalogs presented in Sections \ref{sec:cat1} and \ref{sec:cat2}. During yaw flips and \textnormal{periods when} proton and SXR flux data were not recorded (during orientation changes, instrumental corrections, etc.), we interpolate fluxes from the time before and after the data gap \textnormal{to} attain continuous flux records for the timeline of interest. \textnormal{Streamlined proton and SXR flux data, our resulting catalogs, and time-series visuals are fetched onto a Solar Energetic Particle Prediction Portal\footnote{\url{https://sun.njit.edu/SEP3/index.html}} (SEP$^{3}$).}

\subsection{\textnormal{Catalog I: Solar proton event records}} \label{sec:cat1}
Considering the definition of an SPE as \textnormal{$\geq$10 pfu detections of protons} $\geq$10 MeV, this is the threshold used throughout the process of generating our first catalog. \textnormal{NOAA defines the severity of solar storms using an} S-scale\footnote{\url{www.swpc.noaa.gov/noaa-scales-explanation}} hierarchy of progressively damaging solar events spanning from S1 (minor) to S5 (extreme). \textnormal{An SPE is the baseline for an S1 event (the weakest-ranking solar storm in this hierarchy), primarily affecting high-frequency radio propagation in the polar regions.} Higher scales (S2, S3, S4, \& S5) are associated with events of much higher \textnormal{fluences (cumulative pfus detected during an event)} ($10^2$, $10^3$, $10^4$ \& $10^5$ respectively)\textnormal{, incrementally enhancing both the radiative environment and the damage they cause.}

\begin{figure}
    \centering
        \centering
        \includegraphics[width=0.475\linewidth]{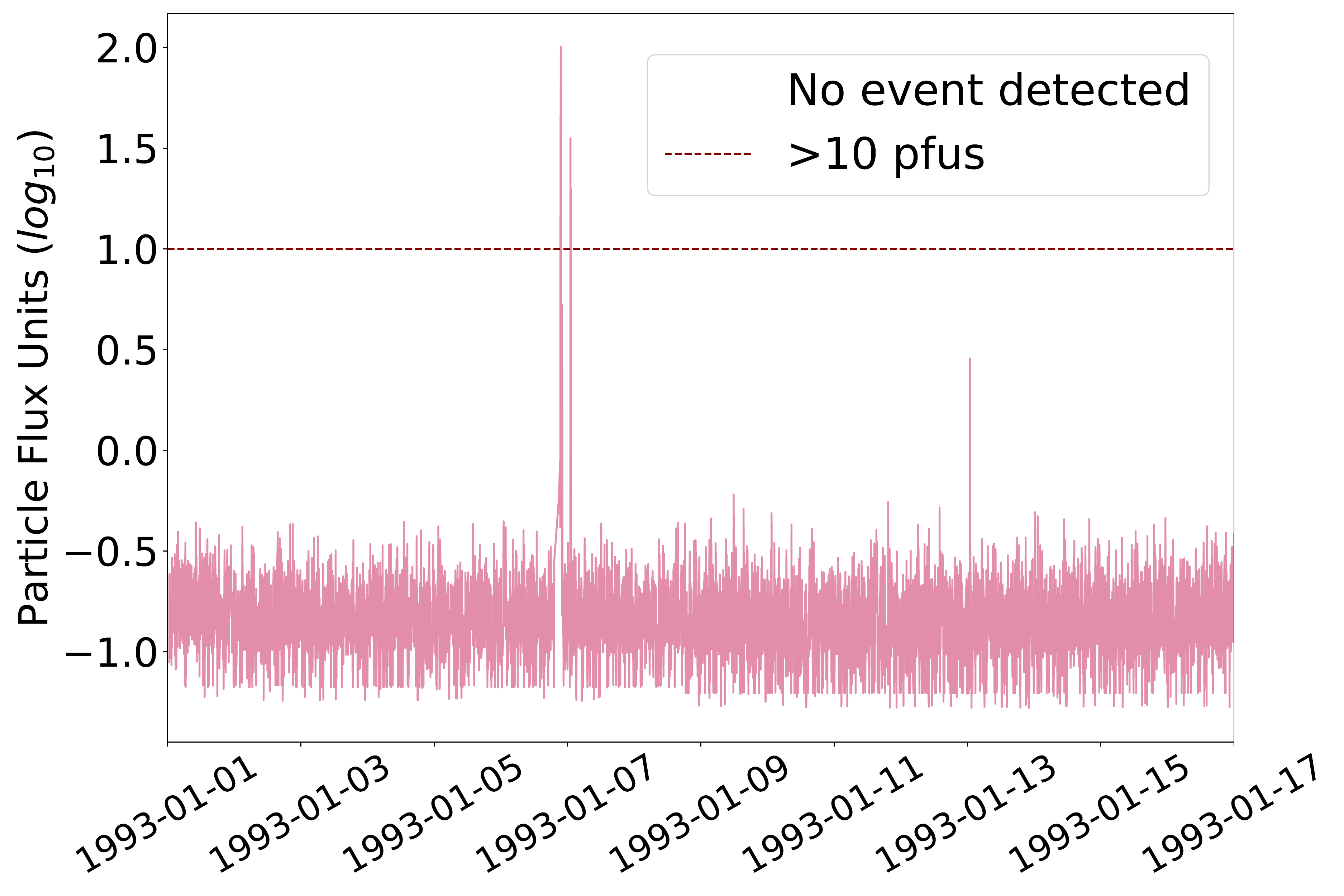}%
        \hfill
        \includegraphics[width=0.475\linewidth]{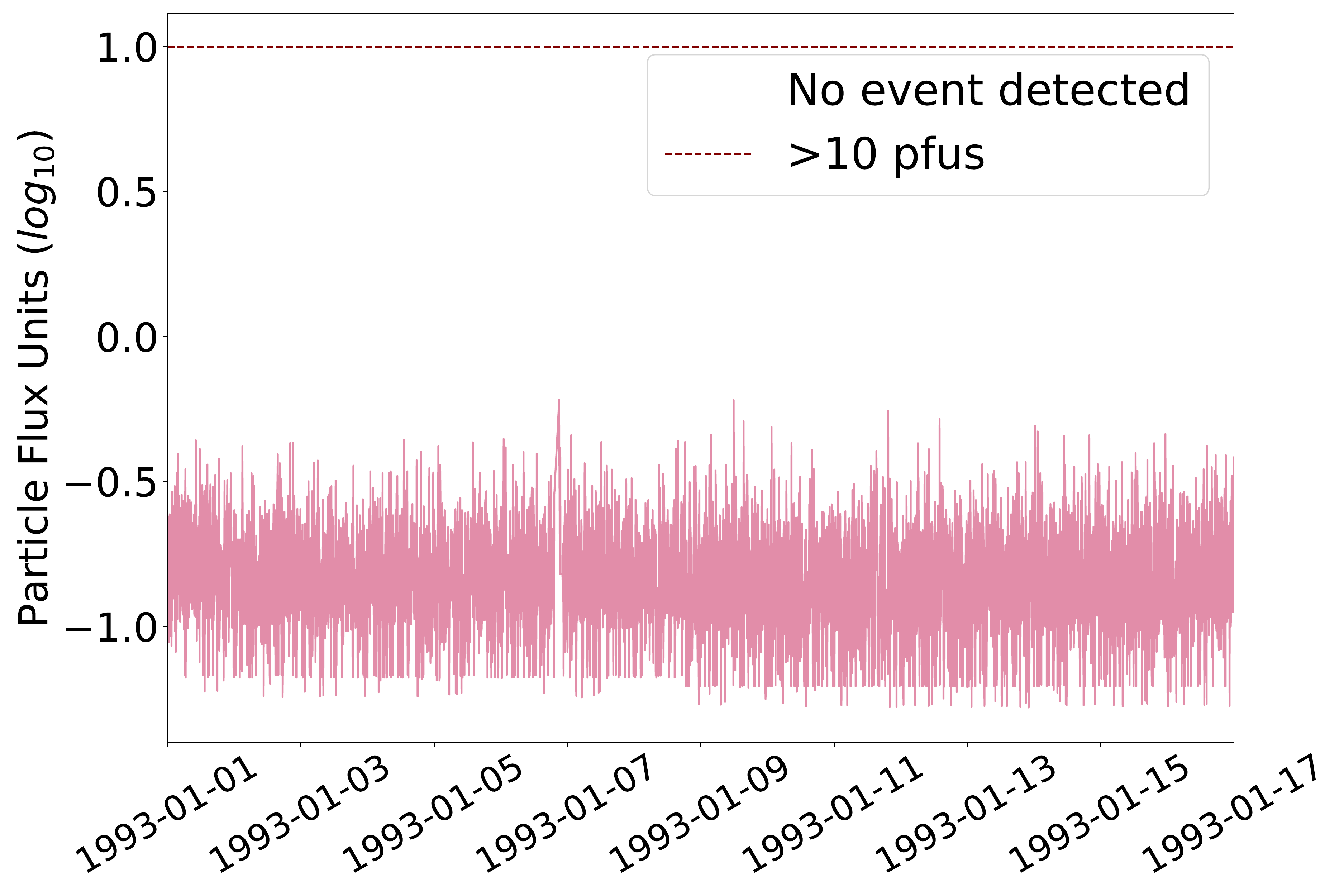}
        \caption{Examples of (left) instrumental effects producing a spike in the data, and (right) how a corrected spike is presented, mirroring a quiet period of the Sun.}
        \label{fig:spike-quiet}
    \vskip\baselineskip
        \centering
        \includegraphics[width=0.475\linewidth]{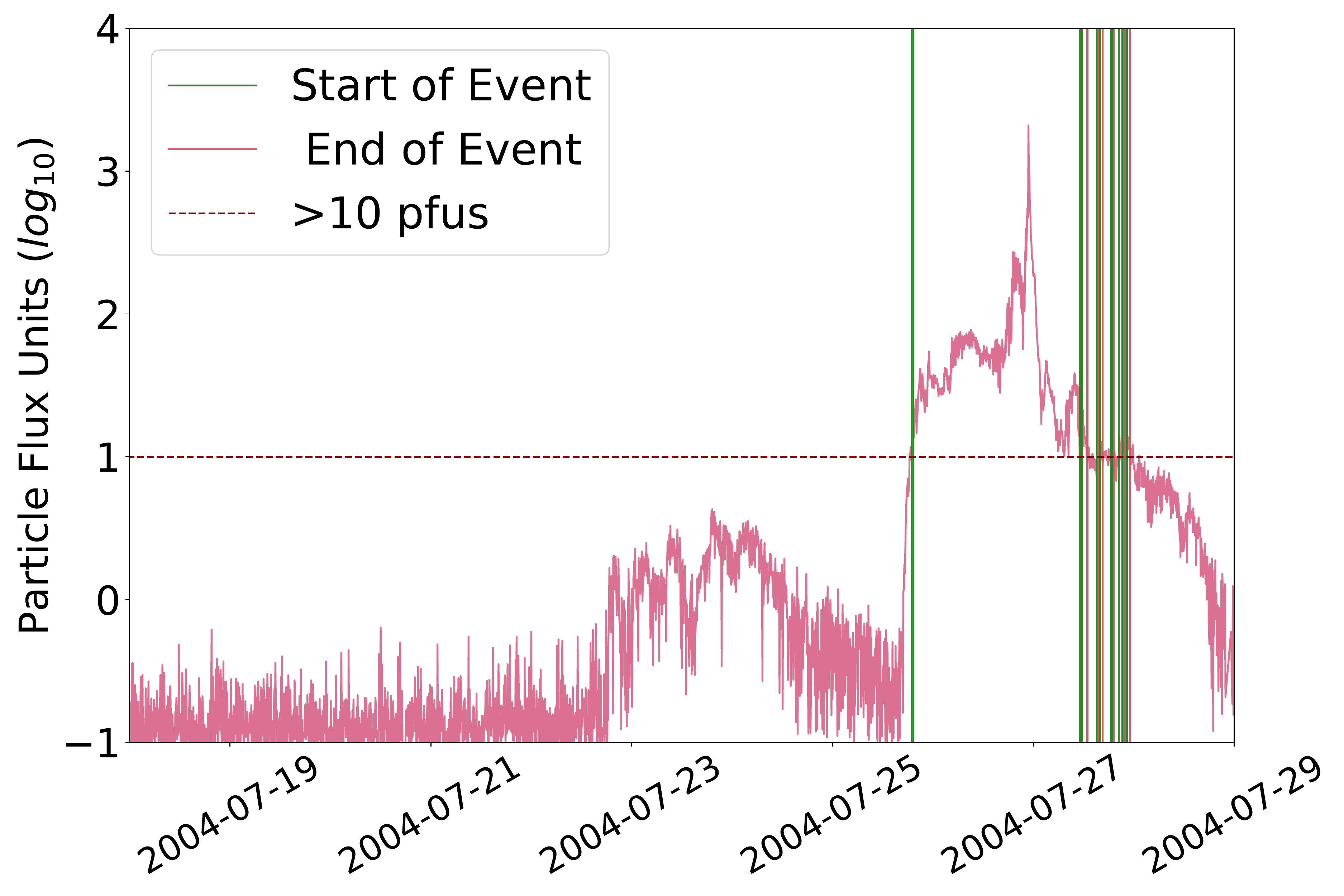}%
        \hfill
        \includegraphics[width=0.475\linewidth]{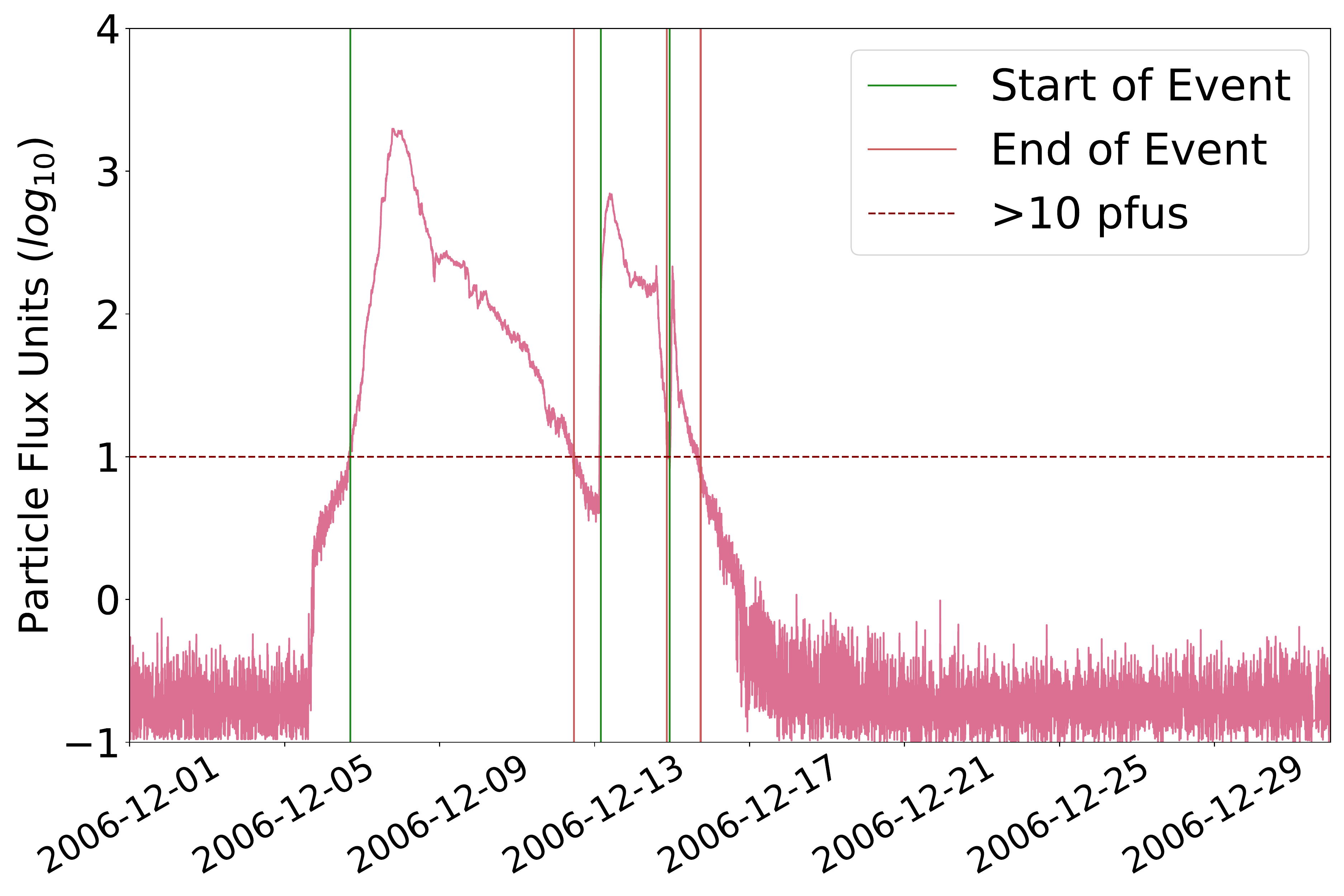}
        \caption{Example of (left) oscillations around the 10 \textnormal{pfu} threshold during an event, and (right) \textnormal{an unambiguous} representation of \textnormal{an} SPE, as they are recorded in our catalog.}
    \label{fig:osc-event}
\end{figure}

To build our catalog of SPE statistics, GOES data \textnormal{were} transformed into a logarithmic form, and cleaned for instrumental effects prior to recording any event parameters- most of which present themselves as \textnormal{spikes} in the data, an example of which is shown in the left panel of Fig. \ref{fig:spike-quiet}. \textnormal{Specifically, we identify spikes where there is a heightened flux value, but the flux right before and after (5 minutes prior and after a spike, given \textnormal{that} GOES proton flux data are provided with a 5-minute cadence) remains an order of magnitude lower. Given the drastic difference in fluxes where we \textnormal{see spikes} without a gradual rise and fall, we are confident that they are instrumental effects, and do not reflect true SEP activity. We correct these by replacing the heightened flux value with the averaged data prior to and immediately after the spike is observed}, as seen in the right panel of Fig. \ref{fig:spike-quiet}. \textnormal{This allows for a more \textnormal{accurate reflection} of the Sun's quiet period while preventing the recording of an SPE when the artificially amplified flux value crosses the $\geq$10 pfu threshold.} Further, given that the threshold can be repeatedly crossed during an event \textnormal{(seen in the }left panel of Fig. \ref{fig:osc-event}), these oscillations are handled by checking if the time between an ongoing event's end and the start of the next is within 30 minutes of each other. Given that oscillations less than half an hour apart are presented in rapid succession and recorded as multiple events over a short period of time, they are instead stitched together as one event in the catalog. These rapid oscillations may represent \textnormal{true proton counts or, and most likely, are instrumental side effects when detecting fluxes.} Opposing these situations, a more prevalent example of \textnormal{an} SPE usually has a more gradual rise and/or fall. The end product of this part of data processing is shown in the right panel of Fig. \ref{fig:osc-event}, reflecting clear start and end dates of \textnormal{an} SPE, barring any undesirable flux representations. \textnormal{At this point, our data set is cleaned and we continue with the remainder of the project.}

\textnormal{\newline In summary, this catalog includes:}

\begin{itemize}
    \item \textnormal{Date \& times for: the start  of an event, an event's peak flux detection, end of an event
    \item Peak event fluxes detected in energy channels: 1 MeV, 5 MeV, 10 MeV, 30 MeV, 50 MeV, 60 MeV, 100 MeV
    \item Fluence (calculated as the integral of detected fluxes) of SPEs in energy channels: 1 MeV, 5 MeV, 10 MeV, 30 MeV, 50 MeV, 60 MeV, 100 MeV}
\end{itemize}
Examples of data products derived from this catalog are shown in \textnormal{Fig.} \ref{fig:daily-stats}.

\renewcommand*{\arraystretch}{1.2} 

\newcolumntype{T}{>{\centering\arraybackslash}m{2.6
cm}} 
\newcolumntype{s}{>{\centering\arraybackslash}m{5.25cm}} 

\begin{figure}
    \centering
    \includegraphics[width=0.8\linewidth]{figures/event-stats.pdf}
    \caption{\centering \textbf{SPE statistics considering proton flux data. We summarize (and include the standard deviation when applicable), the average duration, peak flux, and fluence per event for each SC, among other characteristics. We can see here how much weaker SC 24 activity was compared to SCs 22 \& 23.}}
    \label{fig:daily-stats}
\end{figure} 


\subsection{\textnormal{Catalog II: Daily flux feature statistics}} \label{sec:cat2}
The second catalog produced during this project supplies the input data for our \textnormal{ML-driven} forecasting \textnormal{models}. This consists of numerous features of daily flux data acquired by GOES in both proton and SXR channels. \textnormal{Before model training begins, it is common practice to convert input data feature vectors into a standardized range. An SVM's optimal hyperplane- the boundary between distinct classes, is influenced by the scale of input features, requiring data to be scaled prior to model training. The same is needed for XGBoost models, which are sensitive to the scale of features when trained using gradient-based methods \citep{xgb-norm}. \textnormal{Correspondingly}, we apply logarithmic scaling and minima-maxima normalization to our training sets and scale this transformation to the corresponding test sets. This is primarily done to allow models to differentiate between various patterns and structures in the data without being influenced by each parameter’s intrinsic physical units and dynamic ranges \citep{azim2021}. Namely, the flux features generated in this catalog are (for each- proton, SXR short wavelength, and SXR long wavelength channels; features correspond to the current day’s flux measurements unless otherwise stated}):

\begin{itemize}
\item \textnormal{Added for time-series records only, and not used in the forecast itself: Instrumental data, Dates of observation, \textnormal{GOES} satellite used (with a primary detector when appropriate)
\item Daily aggregated flux data \& statistics: mean, median, minimum, maximum, standard deviation, skewness, kurtosis, and the last measured flux of the previous day.} 
\end{itemize}
\textnormal{In our previous work \citep{slava2021} we applied a 2-hour latency period between the time GOES data are released (10:00 p.m. UTC daily) and the next day (12:00 a.m. UTC). We handled this by applying a 2-hour offset for each day's records, i.e. we +2 hours to every timestamp, so that each day ``ends'' when new GOES data are released. The new data supplies flux records for \textit{only} the next day. In doing this, any predictions we make for the following day are technically done at 10:00 p.m. UTC. We did this to allow 1:1 comparisons between our forecasts and those made by Space Weather Prediction Center at the National Oceanic and Atmospheric Administration \citep[SWPC NOAA; also make daily SPE predictions at 10:00 p.m. UTC, although using different features as input for probability-based predictions,][]{forecaster}. We apply \textnormal{this offset and }consider previous SWPC prediction probabilities as a baseline to assess our ML-driven model performance in Section \ref{sec:results}}.

\subsection{\textnormal{Flux feature importance \& selection}} \label{sec:feature_selection}
As discussed in \citet{bobra} and \citet{slava2021}, \textnormal{including all available features into an ML model} does not necessarily lead to an increase \textnormal{in predictive scores,} and may even result in a notable decrease. In the presence of multiple irrelevant or redundant features, learning methods tend to overfit contributions and become less interpretable or produce entirely inadequate results. A common way to resolve this problem is \textnormal{by implementing feature selection, which works to reduce} supplied data dimensionality by selecting only a subset \textnormal{of the input features} (which in our case, is the catalog discussed in Section \ref{sec:cat2}). \textnormal{We determine each flux feature's ``importance'' by using Gini Importance, Fisher-scoring, and XGBoost (uses an inherent feature importance scheme) to rank each feature's contributions toward reliable \textnormal{forecasts, retaining} those scoring the highest.} This also works to reduce associated computational costs and removes irrelevant features for problems with multi-dimensional data \citep{f-score}.

\textnormal{In order, Gini Importance is computed using a Random Forest structure providing relative rankings of input features indicating how often specific features are selected for node splitting (deciding how to divide data into separate classes). In doing so, Gini Importance quantifies different input features' contributions towards \textnormal{the improvement or decline of} model performance \citep{gini}. Aiming to reduce feature dimensionality like Gini Importance, Fisher-scoring is one of the most popular supervised univariate feature selection methods and is explained in detail by \citet{f-score}. Concisely, Fisher-scoring measures the \textnormal{intra-class variance between features in both positive and negative classes, identifying those that stand out from neighboring features (i.e., those defining the separation between the two classes)}. Lastly, the XGBoost algorithm has a built-in feature importance function. Given that XGBoost is decision-tree-based and forms an ensemble of numerous sub-models (further explained in Section \ref{sec:xgb}), it determines feature importance in a slightly more complicated way compared to Gini and Fisher ranking. Recording how often a feature is used to split a node in the decision trees of the ensemble, the algorithm quantifies each feature’s average contribution to the decision-making process. Summing up these quantities across all trees in the ensemble, XGBoost returns the order of each feature’s contribution toward accurate predictions.}

Comparing feature ranks across all methods, we retain only 9 out of our original 24 features to \textnormal{develop} our prediction models. The ranking of the parameters was completed separately for every solar cycle, yet resulting in the same list of \textnormal{5} top-ranking features \textnormal{(though not in the same order). Fig. \ref{fig:fi} highlights the substantial decline in scores between the leading feature- the previous day's last measured proton flux, compared to the last feature we retain for testing- long SXR flux standard deviation. The remaining features' importance scores become increasingly lower, and we do not consider them from this point on. While only the top 5 features were the same across the three methods, we selected 4 additional features by comparing scores appearing in at least 2 ranking methods (after the initial 5 matches), until scores dropped below $0.02$. We deemed features below this insignificant toward the predictions of SPEs. \textnormal{Figure~\ref{fig:fi} presents these features and their rankings}}. Listed in descending order of their averaged ranking scores, the retained features are the previous day’s last measured proton flux (measured at 10:00 p.m. UTC), proton flux maximum, proton flux standard deviation, short SXR flux mean, proton flux skewness, previous day’s last measured long SXR flux, proton flux median, previous day’s last measured short SXR flux, and long SXR flux standard deviation. It is clear from these retained features that the proton flux value from the previous day dominates in importance compared to the rest of the features in all feature selection measures. This is intuitive given that a rise of SEP intensity on the previous day can easily be used as a predictor that SEPs may cross (or continue crossing, if an SPE is in progress) the 10 pfu threshold the next day. Still, this may not always be the case as shown in the left panel of Fig. \ref{fig:osc-event} where an event may begin abruptly without any indication of rising proton intensity in the days leading up to it.

\begin{figure}
    \centering
    \includegraphics[width=0.75\linewidth]{figures/fi.pdf}
    \caption{\textnormal{A representation of the top 9 ranked features according to Gini Importance, Fisher Scoring, and XGBoost's built-in feature selection methods. Feature ranking order being similar for all SCs of interest, scores are averaged across all \textnormal{cycles and} are shown here. Given that the different methods scale on different ranges, we apply a minima-maxima normalization here \textit{only} for visibility\textnormal{, and present these normalized ranking scores in the legend.} \textnormal{Notably}, the lowest ranked of these features (long SXR flux standard deviation) is barely visible in each chart.}}
    \label{fig:fi}
\end{figure}

With 5 of the 9 finalized features relating to proton flux (3 of which are ranked as the first, second \textit{and} third most important), and 2 relating to each the short and long SXR irradiance channels, we can confidently say \textnormal{that proton flux features are most important towards SPE prediction when using our data set}. This upholds our conclusions from \citet{slava2021} that using SXR features \textit{in addition} to proton fluxes improves performance scores, but that higher scores depend largely on proton flux data.

\subsection{Devising \textnormal{training \& testing data sets} by Solar Cycle} \label{sec:sc_sep}
For the application of ML models, data sets are divided into ``training'' and ``testing'' sets. \textnormal{We do so by splitting our data into the different SCs of interest i.e. SC 22: 1986 -- 1996, SC 23: 1996 -- 2008, SC 24: 2008 -- 2019.} During the training phase, models examine provided data from specific time intervals to learn patterns from and generate predictions for a different time interval \citep{bishop}. \textnormal{Part of the supplied data set is kept blind to the training data, forming the test set. After a model is initialized and parameterized, it is applied to \textnormal{the test} set to generate predictions. These predictions are compared to the true target \textnormal{values to} estimate how well a model may generalize to unforeseen data. For example, if considering training done using SC 23 data, model performance scores are generated based on how accurately the model is able to reproduce the observed events of the specified prediction (testing) window of either SC 22 or 24. All combinations of training and testing sets are used, allowing every SC to serve as the training and testing set at least once (e.g. training done on SC 22 and testing done using SC 23 data, training done on SC 22 and testing done using SC 24 data, and so on).}

\textnormal{Additionally, we stacked training data sets to explore if longer training time intervals allow more precise predictions (i.e. training SC 23 \& 24 together to test with SC 22 data, SC 22 \& 24 to test with SC 23 data, and finally, SC 22 \& 23 to test with SC 24 data). Once these different data sets are defined, we use a minima-maxima normalization with respect to \textit{only} the training set and scale this transformation to the test set. It is important to directly normalize only the training set, otherwise, the model will be ``exposed'' to some of the test set's information and possibly learn from it, giving the model the advantage of having prior knowledge of its target output that it should not- and realistically will not have when applied elsewhere. Once the normalization is appropriately done, we employ SVM and XGBoost algorithms, discussed in Sections \ref{sec:svm} and \ref{sec:xgb}. Because our primary goal is not to parametrize a specific algorithm with minute detail, we use grid searches to modify a few model parameters relating \textit{only} to optimize classification. We then test the effects of oversampling our positive classes using standard oversampling (positive-class duplication), Synthetic Minority Oversampling TEchnique (SMOTE), and Adaptive Synthetic (ADASYN) oversampling, discussed in Section \ref{sec:overgrid}. Model performance in both cases of single-cycle and double-cycle training using SVMs and XGBoost are discussed in Section \ref{sec:results}.}

\textnormal{It is important to note that the data are prepared equally between models to generalize results as much as possible. Concisely, data preparation and model evaluation follow the order: splitting our data set into 3 segments (each SC of interest) to use as training and testing sets, down-selecting input data to 9 flux features contributing most to reliable predictions, normalizing only the training set and scaling this transformation to the testing set, using model-inherent class-balancing parameters or different oversampling methods, performing grid searches to establish optimal parameters to apply to our models, and finally, evaluating model performance considering evaluation metrics \textnormal{discussed in Section \ref{sec:eval}}: True Skill Statistics, Heidke Skill Scores$_{2}$, and recall.}

\section{ML \textnormal{applications to Solar Proton Event forecasting}} \label{sec:apply}
\subsection{Learning methods considered}
\textnormal{Selecting the appropriate model is vital when we consider resolving real-world challenges. Towards our effort of predicting SPEs, we compare the performance of supervised classification using an SVM, and a decision tree-based gradient boosting ensemble algorithm, XGBoost.}

\subsubsection{Support Vector Machines (\textit{SVMs})} \label{sec:svm}
Introduced in 1963, SVMs are supervised classifiers with roots in the theory of statistical learning with the ability to learn from nonlinear decision surfaces with the application of different kernels. \textnormal{Data are assigned to their respective classes depending on where they lie with respect to the hyperplane- features relevant and irrelevant towards the forecasting of SPEs} \citep{bishop}. Black-box machines like \textnormal{SVMs} aim to generate optimal hyperplanes with dimensions mirroring that of the number of features set as input data, determining feature space (the \textit{n}-dimensions where input variables live) \textnormal{dimensionality \citep{bishop}. The} arguments we apply to the \textnormal{SVM,} supplied by the \textit{scikit-learn}\footnote{\url{https://scikit-learn.org/stable/modules/generated/sklearn.svm.SVC.html}} library include a \textnormal{\texttt{radial-basis function} (RBF) kernel, regularization parameter \texttt{C}, a kernel coefficient \texttt{$\gamma$ = scale}, and \texttt{balanced} class weights.}

For SVMs, a kernel refers to a method allowing the application of classifiers to non-linear problems by mapping non-linear data \textnormal{to higher-dimensional space; where data become more easily separable (either linearly, radially, or polynomially, depending on user input). \textnormal{Different kernels available \textnormal{to} use with SVMs allow better data transformations depending on the input data set. A} hyperplane is then built calculating the dot product between the transformed features.} \texttt{Linear} kernels may be considered first given \textnormal{their} low computation needs, allowing \textnormal{quick training and testing capabilities. Oftentimes, other kernels perform better, but it is} simply important to note that this is not always the case. \textnormal{Over all kernels available, the \texttt{RBF} kernel proved to be the better choice when evaluated under parameter-optimization techniques discussed in Section \ref{sec:overgrid}, and is the only kernel employed and discussed from this point on.}

\textnormal{We apply a grid search to the regularization parameter \texttt{C}, which allows an SVM to build a hyperplane with both the largest minimum margin, and one} separating as many instances as possible. \textnormal{The \texttt{C}} parameter decides how to prioritize enhancing the latter. Given that the regularization parameter affects different testing data sets \textnormal{differently}, there is no absolute of whether larger or smaller values lead to more appealing results. \textnormal{The \texttt{$\gamma$ = scale} argument is the \texttt{RBF}} kernel coefficient \textnormal{declaring} how far the influence of a single training data point reaches. Automation by \citet{sklearn} of scaling this parameter to the data alleviates the need to \textnormal{process it manually. Small \texttt{$\gamma$} values consider} data points farther \textnormal{(the minority class) from where most clump together (the majority class)} when defining the separation line. \textnormal{The automated scaling searches between higher and smaller values to achieve the best possible fit for the provided data set, working as almost a proxy to bridge the gap of our imbalance.}

Lastly, \textnormal{we alter the assigned class weights, where the default value is \texttt{None} (all classes are assigned \textnormal{equal importance, i.e. weight = 1)}. Described} in \citet{bishop}, a loss function is a method of evaluating how well an algorithm models its supplied featured data set; i.e. an optimal algorithm minimizes its loss function. The \textnormal{\texttt{balanced} class weight parameter} directly modifies an algorithm’s loss function by varying penalties assigned to classes with different weights. Using this biases the model to favor predictions of the minority class by assigning larger weights to them \citep{tensorflow-balover}. To the same effect, we \textnormal{apply different} oversampling techniques \textnormal{to an SVM, inflating positive-instance relevance in the training data set,} discussed in Section \ref{sec:overgrid}. 

\subsubsection{\textnormal{eXtreme Gradient Boosting; XGBoost}} \label{sec:xgb}

\textnormal{More recently, also exploring SEP prediction and considering TSS and HSS$_{2}$ as metrics to evaluate model performance, \citet{patrick} show how ensembles lead to better and more robust (in terms of experiment-to-experiment variations) predictions. With increasing advances focusing on deep learning techniques, \citet{xgboost-intro} found the open-source gradient boosting algorithm XGBoost developed in 2016 by \citet{xgboost-2016} to be more effective for predictions compared to multiple more complicated deep-learning models. Therefore in addition to an SVM, we test the performance of XGBoost when building SPE forecasts using our data set. The algorithm represents a boosting ensemble classifier and uses a \textit{gradient descent} framework, generating new models from the output of preliminary models. As an iterative decision tree-based learning algorithm, one ends up with an ``ensemble'' of sub-models working to optimize each new learner. This process terminates once the optimal model (as the loss function is minimized as far as possible,) is reached. Building this ensemble allows accounting for multiple modeling results, leading to more stable and generalized results.} We implement this method, again using the \textit{scikit-learn} library, with the wrapper class \texttt{XGBClassifier}. Reiterating that our primary goal is not to parametrize a specific algorithm with minute detail, we specify only two default parameters: \texttt{booster : gbtree} and \texttt{scale positive} weights. Respectively, the \texttt{gbtree} specification indicates using tree-based models to incrementally build an ensemble. Other options include building an ensemble while dropping a sub-model per iteration, or using linear functions instead. Lastly and similar to the SVM's \texttt{balanced} class weight, XGBoost has an intrinsic parameter to \texttt{scale positive} weights, balancing each data class instead of leaving them widely imbalanced. With the original imbalances $\sim \frac{\#~of~positive~instances}{\#~of~negative~instances} << 1$ for each SC of interest, \texttt{scale positive} weights work to alter weights applied to each data class and point to bring this closer to 1.

\subsection{\textnormal{Considering oversampling techniques \& grid search optimization}} \label{sec:overgrid}
\textnormal{In lieu of our original imbalance \textnormal{(Fig. \ref{fig:imbalance})}, oversampling calls for synthetically repeated observations of the minority class (days with SPEs), until its frequency in the data set is comparable to that of the majority class (days without SPEs). Doing this brings equal representation of both classes, as well as an artificially extended data set. Following the classical definition of oversampling, we use standard techniques to inflate the positive cases of days with SPEs observed on the multiplicity of reaching an imbalance ratio $\frac{\#~of~positive~instances}{\#~of~negative~instances}\sim$1. While this seems analogous to the \texttt{balanced \& scale positive} weight parameters inherent to SVM and XGBoost classifiers, oversampling alters the training data directly instead of adjusting class weights, providing different results due to their intrinsic methodologies. Doing this allows us to evaluate how SVM and XGBoost models perform when considering a data set with two classes of approximately the same weights instead of one drastically imbalanced (original imbalance ratios follow: $\frac{\#~of~positive~instances}{\#~of~negative~instances}\sim$ 0.053, 0.058, and 0.032 for SCs 22, 23, \& 24 respectively).}

\textnormal{The recent work of \citet{new-smote} utilizes SMOTE to account for the maximum-likelihood estimation (MLE) for SPE prediction. They include SXR, radio fluence, and flare helio-longitudes as input features to take into account particle propagation from the Sun. Paired with a synthetically oversampled data set, they found improved probabilities of detection compared to basic MLE and weighted MLE, with predictive scores increasing from 0.76 and 0.75 (respectively) to 0.80. Because of this, we explore how this technique may \textnormal{enhance our models}. SMOTE works via linear interpolation, replicating minority instances between pre-existing positive data points to increase their presence in the data set. SMOTE uses an intrinsic k-nearest neighbor method (where we define k=5) to select points for interpolation. Another oversampling technique we use, ADASYN, works similarly to SMOTE, except it focuses on generating data points in regions where the class imbalance is most prominent, giving more importance to positive instances harder to ``reach'' and learn from, reducing the risk of overfitting. ADASYN uses the k-nearest neighbor method on \textit{each} data point so that each minority class' data points are associated with different neighborhoods. This allows for a more complicated data set to be generated, making the learning phase more complex (and realistic). Both SMOTE and ADASYN oversampling ensue until the ratio $\frac{\#~of~positive~instances}{\#~of~negative~instances}\simeq{}1$.}

\begin{figure}[ht!]
    \centering
    \includegraphics[width=.75\linewidth]{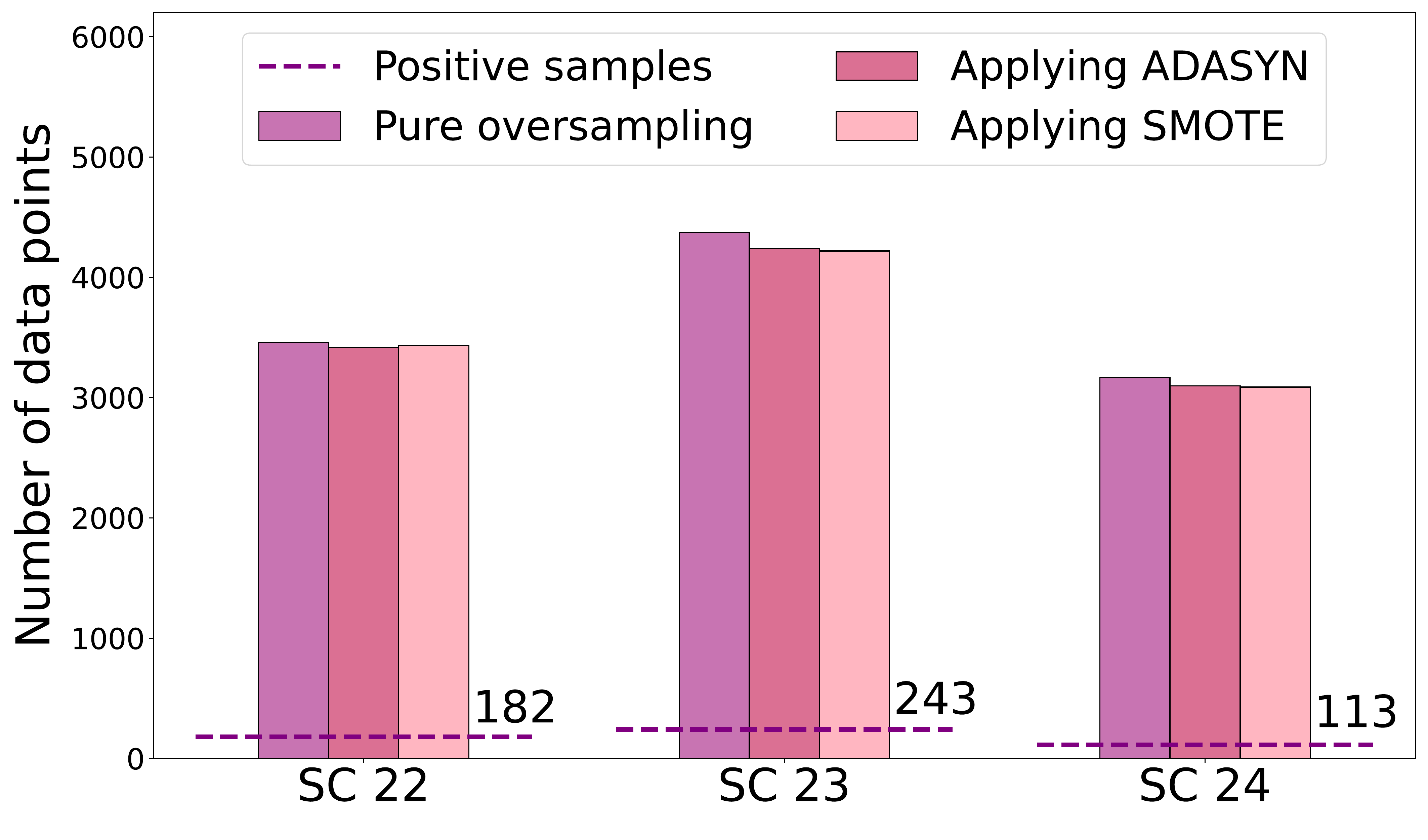}
    \caption{\textnormal{Differences between the number of data points after oversampling using standard (positive-class duplication), SMOTE, and ADASYN are shown here. Albeit on the order of tens, it is worth noting that the different methods intrinsically handle data differently.}}
    \label{fig:over-diff}
\end{figure}

\textnormal{As a final imbalance-handling technique, we test the effects of standard oversampling on predictive models. Here, we simply duplicate the positive cases in our target training set as many times as needed until these cases appear just as frequently as the negative cases in our input catalog. In doing so, the training set is inflated to have as many positive cases as negative cases, but we apply none of the intricacies associated with SMOTE and ADASYN. The (small) differences in the number of positive cases after oversampling using these techniques are shown in Fig. \ref{fig:over-diff}. Applying each technique to the input for an SVM and XGBoost algorithm, the synthesized data replaces the original training set, and the models continue working towards making predictions.} 

Our final modification to each model is using a parameter optimization technique to alter the most basic classification parameters of our models. Grid search is a process capable of automatically parsing through a specified range of numbers or strings appropriate for model arguments to find optimal values for evaluation, and \citep[we use the module provided by][]{sklearn}. \textnormal{This} pinpoints \textnormal{parameter values }leading to the most accurate predictions. The grid search was \textnormal{performed when initializing the prediction models during the training phase, optimizing parameters to reach the maximum possible True Skill Statistic scores (TSS, see Section~\ref{sec:eval}).}

\textnormal{We contain our parameter grid search on only 2 parameters for an SVM,  kernels- between \texttt{linear, polynomial, and radial basis functions}, finding \texttt{RBF} to be the optimal choice, and the regularization parameter \texttt{C}. Interestingly, changes in the \texttt{C} parameter did not lead to significant changes considering the cross-validation TSS of each model when parsing through a wide range of ${2}^{1}-{2}^{10}$, which \citet{best_param_c} determined to be satisfactory.}

\textnormal{We only apply one parameter of XGBoost to a grid search; the \texttt{learning rate}. The default value for this, defined by \citet{xgb-doc} is 0.3 (on a range of 0-1) to help prevent overfitting. When building an ensemble, the learning rate inherently decreases as the weights of each feature change at every ``boosting'' step. This allows the process to not ``learn'' too much from previous steps, which would otherwise incrementally build upon the initial feature weights to falsely make the first-built ensemble the most optimal. We see more variation in model accuracy here compared to how changes in \texttt{C} altered SVM performance.}

\textnormal{Altering minimal hyperparameters, we are not fine-tuning the SVM or XGBoost algorithms to fit our specific training and testing configurations. Our results using grid searches when compared to default-parameter model evaluations are just slightly better (on the order of $\sim{10}^{-2}$) and therefore, are the only results considered from this point on (e.g. when discussing the default SVM model, we regard the SVM model with its default parameters \textit{with} a grid search applied on the model-appropriate arguments mentioned above). We use grid searches on each type of model evaluation: each model (SVM and XGBoost) with its default parameters, each model with imbalance-handling weights applied (\texttt{balanced} or \texttt{scale positive}), and the model applying standard / SMOTE / ADASYN oversampling (individually) on the training set. We assess each model's performance when making predictions based on short (trained on a single SC) and long (trained on two SCs) timescales.}

\subsection{Model evaluation metrics} \label{sec:eval}

\begin{figure}[b]
    \centering 
        \centering 
        \includegraphics[width=0.9\linewidth]{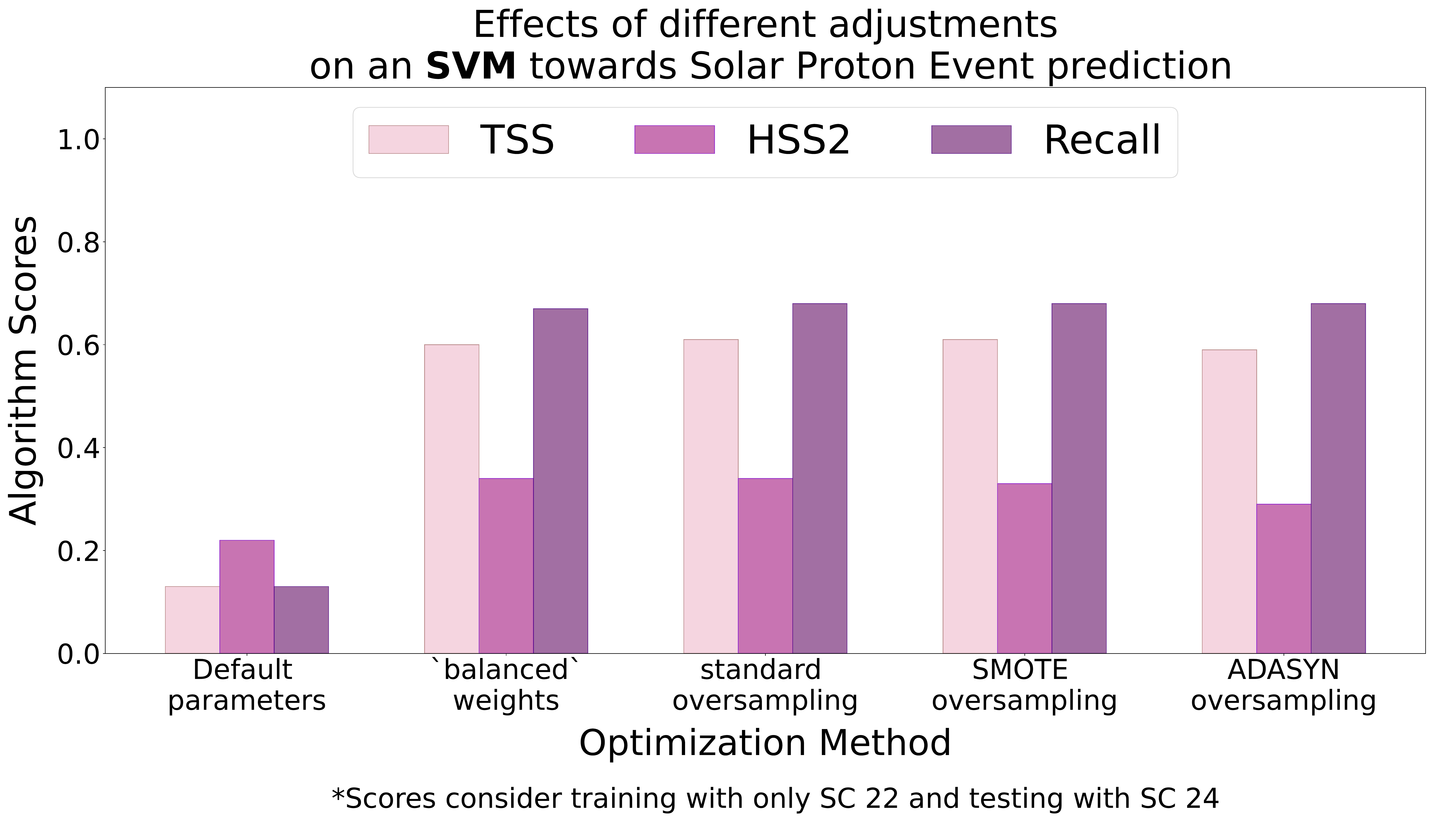}%
        \caption{\centering \textnormal{Resulting TSS, ${HSS}_{2}$, and recall scores of various adjustments made to single-cycle training data on an SVM.}}
        \label{fig:svm-1s}
\end{figure}

\begin{figure}[ht]
    \centering 
        \centering 
        \includegraphics[width=0.9\linewidth]{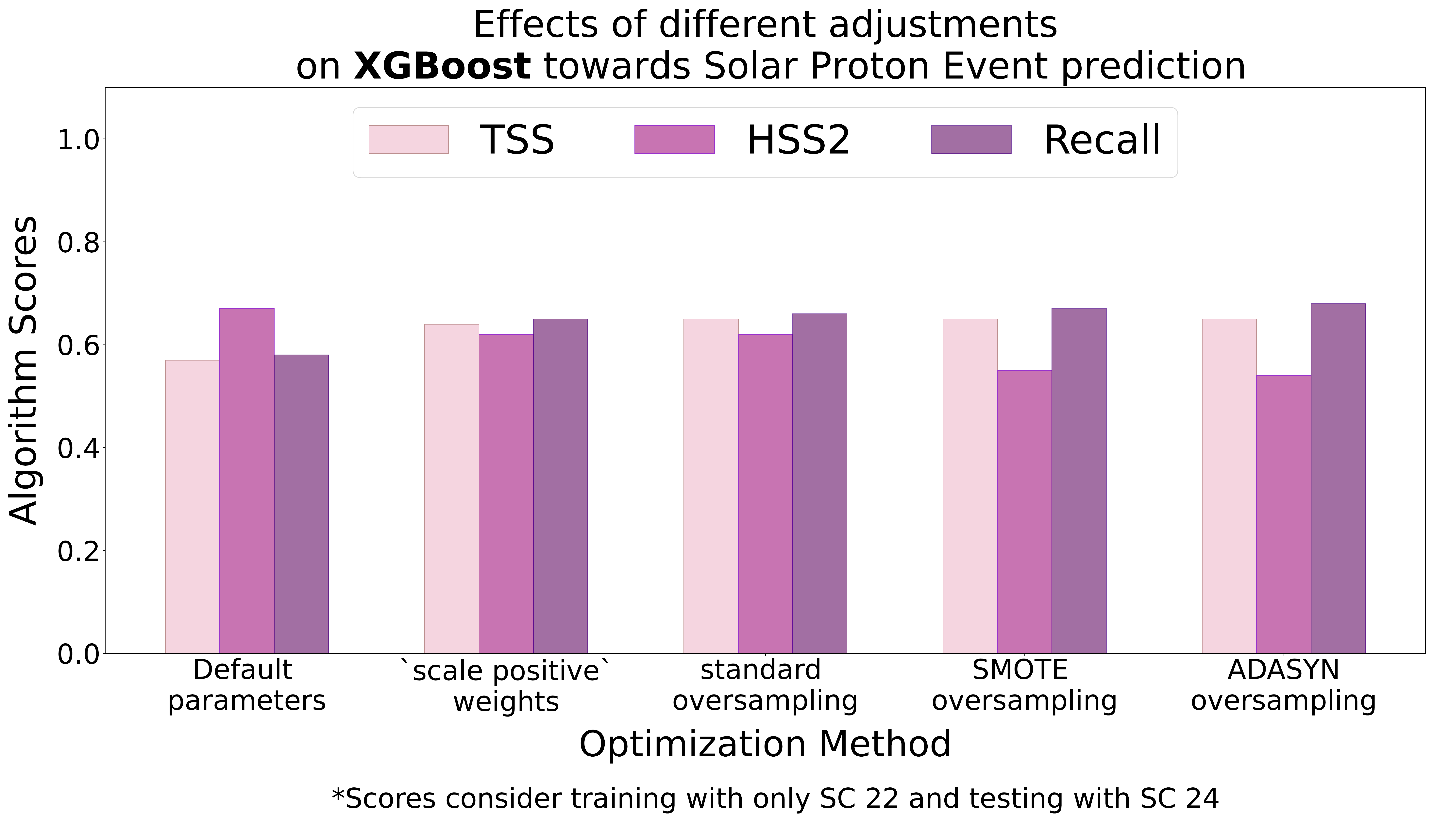}%
        \caption{\centering \textnormal{Resulting TSS, ${HSS}_{2}$, and recall scores of various adjustments made to single-cycle training data on an XGBoost model.}}
        \label{fig:xgb-1s}
\end{figure}

\textnormal{Predictive scores and output may vary each time an algorithm is run, calling for methods to provide an average performance assessment over a large number of model iterations. k-fold cross-validation is one of the most popular methods to do so. The term \textit{k-fold} refers to how the entire training data set is partitioned into multiple subsets of equal sizes, or folds. k = 10 (10 folds) is commonly used \citep[e.g.,][]{kfold-ex1,kfold-ex2,kfold-ex3}, and \citet{why-10} discusses how on average, a model trained on 10 folds can be considered closest to that most effectively reducing prediction errors. After specifying the number \textnormal{of folds (which we leave as the default k=10)} to partition the data \textnormal{into,} the model is trained on (k-1) folds, leaving the last for testing (each fold is used as a test set only once). The final performance score assigned to the model is aggregated from performance metrics across all train-test splits, to provide a more comprehensive evaluation of predictive power. CV optimizes model performance with respect to a specified scoring scheme, for which we use TSS. As opposed to the default k-fold technique, we implement \textit{scikit-learn’s} modification of this, the \textit{stratified k-fold} method. The difference here is that each fold preserves the number of positive instances in the data set, allowing positive instances to appear as many times as possible during the training phase to enhance model performance. Examples of these cross-validated results using single-SC training are shown in Figs. \ref{fig:svm-1s} \& \ref{fig:xgb-1s}. We can see how consistent XGBoost scores remain throughout different training and testing combinations and oversampling methods, while SVMs show more variation \textit{along with} reduced performance scores.}

\textnormal{Evaluation metrics TSS and $HSS_{2}$ are widely used in space weather forecasting \citep{patrick}. Following SWPC's formulation \citep{hss-meaning}, these are defined as:}
 
\begin{equation} \label{tss}
TSS = \frac{(TP)}{(TP)+(FN)} - \frac{(FP)}{(FP)+(TN)},
\end{equation}

\begin{equation} \label{hss2}
HSS_{2} = \frac{\textnormal{2 $\cdot$ (TP$\times{}$TN --- FN$\times{}$FP)}}{\textnormal{(TP+FN)$\times{}$(FN+TN) + (TP+FP)$\times{}$(TN+FP)}}, 
\end{equation} \vskip.15in 

\textnormal{\noindent where \textit{TP} = true positive, \textit{FN} = false negative, \textit{FP} = false positive, and \textit{TN} = true negative forecasts. \citet{mason2010} discuss how HSS$_{2}$ measures the performance of the forecasting model concerning random chance forecasts. TSS ranges from -1 to +1, where +1 indicates predictions made in perfect agreement with the testing set. Any misclassification then reduces this score accordingly. Values $\leq 0$ indicate model performance no better than a purely random forecast \citep{azim2021}. HSS$_{2}$ ranges from -1 to +1 as well, where an algorithm of complete accuracy obtains a score of +1, an algorithm forecasting no events obtains a score of 0 \citep{hss-meaning}, and an algorithm no better than random guesses obtain a negative score. An advantage of using TSS to validate our algorithm is its neutrality towards the class-imbalance ratio- the score itself does not depend on the inequality in trials. \citet{tss-good1} and \citet{tss-good2} echo this in their respective works, agreeing that TSS is an adequate measure of the overall classifier quality and should be the standard to use in comparisons of the performance of various classifiers for flare, weather, and rare-event forecasting \citep{azim2021}. Therefore in this work, we optimize model performance using cross-validation with respect to TSS.}


\textnormal{After the models undergo the process of \textnormal{10-fold} CV, different performance measures are generated. A caveat here is that given the imbalance, CV scores falsely show model accuracy to be $\sim$ 97\% even when \textnormal{failing to predict} numerous SPEs. Accuracy is therefore not a reliable measure of predictive power, and we use ``recall'' to describe model capabilities instead. Recall is the calculation of a model's ability to predict/observe all positive instances supplied by the test set. Working directly on the ratio of predicted SPEs to the total number of observed SPEs, it is an unambiguous metric to evaluate how well the model has been trained to reproduce observations. Recall scores lie in a 0 to 1 range, where 0 reflects the model's inability to identify \textit{any} positive instances, and 1 means that the model correctly identified \textit{all} positive instances of the test set.}

\begin{figure}
    \centering
    \includegraphics[width=.99\linewidth]{figures/final-scores.pdf}
    \caption{\centering SVM \& XGBoost performance scores for each evaluated training-testing configuration. The shaded cells are the \textit{only} instances where an SVM produces better \textbf{(or comparable) results than XGBoost. Column 1 labels the different training-testing configurations referred to in Figures \ref{fig:tss-comp} \& \ref{fig:over-meth}. Also note that each ``model adjustment'' here includes the grid search optimization discussed in Section \ref{sec:overgrid}. \textit{$^{*}$Inherent class weight balancing adjustments are the \texttt{balanced} and \texttt{scale positive} weight parameters inherent to SVM \& XGBoost respectively.}}}
    \label{fig:scores-use} 
\end{figure} 

\section{\textnormal{Solar Proton Event prediction results \& discussion}} \label{sec:disc}

\textnormal{Understanding relations between physical parameters of solar radiation and energetic particle fluxes across different SCs is not an extensively studied problem for the prediction of SPEs. Following our conclusions in \citet{slava2021} and using exclusively operational proton and SXR GOES flux data, our statistical catalog (model input data set; Section \ref{sec:cat2}) records a total of 24 flux features. We reduce this to the \textit{top 9 ranked} features based on results of Gini Importance, Fisher scoring \& XGBoost feature selection. By training SVM and XGBoost models using only these features along with the aforementioned grid search-applied parameters and oversampling methods, we obtain the performance metrics (TSS, ${HSS}_{2}$, and recall) shown in \textnormal{Fig.}~\ref{fig:scores-use}.}

\subsection{Resulting \textnormal{forecast accuracy}} \label{sec:results}
We find that the SVM model performance varies for single-SC training and testing runs across the three SCs, as well as across the implemented oversampling strategies, reflected by the large spread of scores: TSS ranging between 0.12 - 0.68, HSS$_{2}$ between 0.19 - 0.50, and recall between 0.12 - 0.88. The lowest of these scores are associated with SVMs used with default parameters and with no class-imbalance treatment strategy applied. This reveals that without balancing the training data set, the \textnormal{classifier demonstrates} inadequate results. Interestingly and in contrast, the XGBoost algorithm varies only about half as much, with TSS ranging between 0.56 - 0.75, and recall between 0.56 - 0.90, but a somewhat larger difference with HSS$_{2}$ between 0.22 - 0.75. For XGBoost, class-imbalance treatments typically result in higher TSS and recall scores, but simultaneously reduce HSS$_{2}$. Compared to SVMs used with default parameters, XGBoost used with default parameters performs significantly better, with performance metrics being $\sim$3x higher than those obtained with a default SVM. The shaded cells in \textnormal{Fig.}~\ref{fig:scores-use} indicate \textnormal{training-testing cases} where an SVM performed comparably to, or better than XGBoost \textnormal{in terms of the different metrics}. As one can see, there are only a few shaded cells (only 23 out of 135), demonstrating the overall enhanced performance of XGBoost.

\textnormal{Further analyzing these scores, we summarize the highest prediction scores and identify median TSS scores in \textnormal{Fig.} \ref{fig:tss-comp} for every class-imbalance treatment technique (weight adjustments or oversampling). We again observe that XGBoost consistently sees improved median TSS scores compared to SVMs by $\sim$0.04 - 0.11 (columns 5 \& 8). The peak TSS scores (columns 4 \& 7) demonstrate approximately the same range of improved scores. This allows us to conclude that XGBoost is performing statistically better for the considered problem compared to an SVM. \textnormal{Fig.} \ref{fig:over-meth} presents the highest TSS, HSS$_{2}$, and recall scores found for single-SC and double-SC training timescales, and the corresponding oversampling techniques. While the oversampling strategy differs, the strongest HSS$_{2}$ and recall scores are again higher for the XGBoost model, except for \textnormal{the} experiment presented in the last row. Interestingly, this \textnormal{figure} shows that ADASYN more often leads to the maxima scores in comparison with other techniques (for both SVM and XGBoost).}

\textnormal{Figures ~\ref{fig:tss-comp}~and~\ref{fig:over-meth} help understand which class-imbalance treatment (i.e., various oversampling methods or class weight adjustments) leads to the best model performance. Columns 5 \& 8 in \textnormal{Fig.} ~\ref{fig:tss-comp} show that all these techniques generate scores very close to each other in terms of their median TSS scores. The only significant difference occurs for XGBoost-based models trained on double-SC time intervals, where standard oversampling and class-weight adjustment perform notably worse (TSS = 0.66$\pm$0.03 and TSS = 0.67$\pm$0.03 respectively) when compared to SMOTE (TSS = 0.72$\pm$0.02) and ADASYN (TSS = 0.70$\pm$0.02). Overall, we can generalize the above findings by stating that the employment of \textit{any} balancing technique considered in this work improves predictions, with no clear preference for a single technique.}

\begin{figure}
    \centering
    \includegraphics[width=0.9\linewidth]{figures/tss-comp.pdf}
    \caption{\centering \textbf{Oversampling techniques used with both (a) an SVM and (b) XGBoost, showing the maximum TSS obtained from each method across different timescales. Median absolute deviations are also shown.}}
    \label{fig:tss-comp}
\end{figure} 

\begin{figure}
    \centering
    \includegraphics[width=0.9\linewidth]{figures/over-meth.pdf}
    \caption{\centering \textbf{Maximum TSS, ${HSS}_{2}$ and recall obtained using (a) an SVM and (b) XGBoost across different training timescales. Columns 5 \& 8 show the oversampling technique used on the training data to achieve these maximum scores. \textit{$^{*}$Note: n/a applies to default model parameters, with no weight or positive-class adjustments.}}}
    \label{fig:over-meth}
\end{figure}

\textnormal{The data flows utilized for SPE prediction in this work (proton and SXR fluxes) are available in \textnormal{real time}. Therefore, it is possible and meaningful to compare the effort in this paper with historical daily operational predictions of SPEs. Here, we examine the performance of developed ML models \textnormal{considering} previous SC (SCs 22 -- 24) operational SWPC forecasts of SPEs. The SWPC NOAA forecasts considered in this work are daily probabilistic forecasts (in contrast with binary predictions made by ML models), with probabilities ranging between 1 and 99 on the possibility of an SPE occurring the next day. \citet{forecaster} describe how these forecasts relied on corrections made manually based on forecaster experience. The forecasts made by SWPC consider predictions made for three consecutive days, but we only use next-day predictions for a direct comparison between these forecasts with SVM and XGBoost models. To convert the probabilistic forecast to binary forecasts, we find the probability thresholds (i.e., the minimum probability starting from which a positive prediction is issued) that lead to the highest TSS or HSS$_{2}$ on the training data set, and apply it to the test data set. In addition, we analyze a persistence model as a baseline to assess our prediction scores. This model is straightforward and not ML-based~--- it does not train and test on data to make predictions. The persistence model uses the previous day's input of whether or not there was an SPE observed. If there was, the model makes a positive prediction for the next day; if not, the model predicts no events the next day. We show these scores in \textnormal{Fig.} \ref{fig:pm_swpc}, and compare them with the \textit{highest} SVM \& XGBoost performances \textnormal{(in terms of TSS)} obtained from double-SC trained predictions.}

\textnormal{There are several observations to mention based on \textnormal{Fig.}~\ref{fig:pm_swpc}. First, the resulting metrics from the persistence model are very stable, ranging between 0.65 - 0.70 considering all three metrics. This model also demonstrates, on average, the highest HSS$_{2}$ of 0.65 - 0.68 across the considered models. Here we want to mention that XGBoost with default parameters (the scores are presented in \textnormal{Fig.}~\ref{fig:over-meth}) leads to higher HSS$_{2}$ (0.74 and 0.75 for single-SC and double-SC training intervals respectively) than that found with the persistence model. This default model also led to acceptable TSS scores of 0.64 and 0.65 for the same experiments. Interestingly, the probabilistic SWPC NOAA results show significant variations from solar cycle to solar cycle. While the scores for SC 22 events were TSS = 0.49 and HSS$_{2}$ = 0.20~--- significantly lower with respect to persistence forecasts, scores increased to TSS = 0.69 and HSS$_{2}$ = 0.59 for the next two cycles, slightly outperforming the persistence forecast in terms of TSS scores.}

\begin{figure}
    \centering
    \includegraphics[width=0.9\linewidth]{figures/pm_swpc.pdf}
    \caption{\centering \textbf{Comparing SVM \& XGBoost performance to NOAA SWPC’s previously predicted SPE probabilities and a persistence model. For SVM \& XGBoost, we show scores obtained when training each model on the remaining two SCs when considering each testing cycle. \textit{Note, the persistence model does not follow a training-testing framework.}}}
    \label{fig:pm_swpc}
\end{figure}

\textnormal{It is also clear from \textnormal{Fig.}~\ref{fig:pm_swpc} that XGBoost typically performs better when considering TSS and recall metrics, albeit with the slightly lower TSS = \textnormal{0.68 compared to} SWPC NOAA forecasts during SC 24 (TSS = 0.69). In comparison, SVM and SWPC probabilistic forecasts typically show weaker performance in any training-testing case~--- together, these models account for the highest variance across all metrics, as well as the lowest measured TSS and ${HSS}_{2}$. Lastly, the persistence model showed the lowest variance in every metric, the highest resulting ${HSS}_{2}$, and the cost of the typically lower recall. It is worth noting that the higher ${HSS}_{2}$ are associated with the non-ML-driven models- this remains true even when the ML models are optimized with respect to ${HSS}_{2}.$. As discussed in Section \ref{sec:eval}, increased ${HSS}_{2}$  indicate that the evaluated model performs significantly better than random guessing. We can account for these higher scores reflecting ``less randomness'' in SWPC forecasts given that it receives external input (from an experienced forecaster) that would only work to make predictions more accurate \citep{forecaster}. Including relevant features in the model in this way could therefore enhance predictive abilities. Human input would also be very beneficial in cases where data needs to be re-assessed to continue making predictions because a machine alone may not know when/how to correct retrieved data. With persistence models, we already have knowledge of the occurrence of an SPE, and the model only continues its last prediction. With SCs dominated by \textnormal{non-eruptive periods,} the relatively stable environment is reflected in the data and we see no random changes other than changes in labels from `0' to `1' and vice versa. This model's input can be very similar to the ``previous day's last measured proton flux'' feature (Section \ref{sec:feature_selection}). Given the small model input, predictions align well with the observed outcomes, and we see improved ${HSS}_{2}$ compared to the ML models. However, in the ML-based models we use, we have 8 additional flux features contributing to the accuracy of future forecasts. These other features allow the machines to learn new data and update forecasts when needed, building patterns between data leading up to SPEs rather than simply checking SEP counts from the day prior. Overall, it is worth noting that the ML-driven models, if tuned properly, outperform both the persistence model and SWPC NOAA operational forecasts across all three SCs while being based on operational data flows.}

\subsection{Assessing SPE-predictive model cross-transferability} \label{sec:cyc-to-cyc}


\textnormal{One of the interesting questions we analyze in this work is how model performance depends on the solar cycles on which they are trained/tested. SPEs remain relatively rare, leading to slightly different statistical properties and number of days with events \textnormal{during} different SCs (see \textnormal{Fig.}~\ref{fig:daily-stats}). Interestingly, the differences in the forecast performances are not so drastic. \textnormal{Fig.}~\ref{fig:tss-comp} illustrates that the median TSS scores for single-SC and double-SC training are generally comparable, with the median absolute deviations not exceeding 0.03. Yet, the individual experiments may demonstrate significant variations. For example, let us consider SMOTE oversampling for the XGBoost classifier in \textnormal{Fig.}~\ref{fig:scores-use}. Below is a summary of TSS scores, with the training data intervals indicated in parentheses:
\begin{itemize}
    \item Predictions for SC 22: TSS = 0.72 (SC 23), TSS = 0.68 (SC 24), TSS = 0.72 (SC 23 \& 24);
    \item Predictions for SC 23: TSS = 0.75 (SC 22), TSS = 0.71 (SC 24), TSS = 0.75 (SC 22 \& 24);
    \item Predictions for SC 24: TSS = 0.65 (SC 22), TSS = 0.66 (SC 23), TSS = 0.64 (SC 22 \& 23);
\end{itemize}
We show the same for HSS$_{2}$:
\begin{itemize}
    \item Predictions for SC 22: HSS$_{2}$ = 0.53 (SC 23), HSS$_{2}$ = 0.26 (SC 24), HSS$_{2}$ = 0.42 (SC 23 \& 24);
    \item Predictions for SC 23: HSS$_{2}$ = 0.60 (SC 22), HSS$_{2}$ = 0.32 (SC 24), HSS$_{2}$ = 0.52 (SC 22 \& 24);
    \item Predictions for SC 24: HSS$_{2}$ = 0.55 (SC 22), HSS$_{2}$ = 0.59 (SC 23), HSS$_{2}$ = 0.56 (SC 22 \& 23);
\end{itemize}}

\textnormal{We can make several observations following these scores. First, the TSS results were typically smaller for SC 24, irrelevant to the training interval used (SC 22 or 23). For example, median TSS scores across all the considered training intervals were TSS = 0.72 for SC 22, TSS = 0.75 for SC 23, and TSS = 0.65 for SC 24. Second, \textnormal{single-SC training with} SC 24 typically led to \textit{less accurate} forecasting models (in terms of TSS and HSS$_{2}$) \textnormal{when training with} SCs 22 or 23. Interestingly, for double-SC training, the inclusion of SC 24 does not lead to an increase in TSS scores and even leads to a decrease in HSS$_{2}$. Further, according to \textnormal{Fig.}~\ref{fig:daily-stats}, the median properties of SPEs during SC 24 were comparable to the properties of SPEs for SC 22, so we cannot claim that the population of SPEs for SC 24 was statistically different from that of SC 22. We also note that this effect was not observed in SVMs trained on SC 24; the TSS and HSS$_{2}$ were comparable to the alternatives, yet still \textnormal{quantitatively} smaller than in XGBoost experiments.}

\textnormal{One fundamental difference between SCs 22, 23, and 24 is in the number of SPE events (or days with enhanced proton flux). \textnormal{Fig.}~\ref{fig:daily-stats} points out that there were just 113 days during SC 24 when proton fluxes were enhanced, whereas SCs 22 \& 23 saw 182 and 243 of such days, respectively. Overfitting is a known and expected side-effect given the nature of learning models \citep{xgb-over}, and the problem of how rare SPEs are comes into play here, as the most common suggestion to reduce overfitting is by collecting more samples to provide models more data points to learn and generate complex patterns from. It is also documented that synthetic oversampling methods contribute to model overfitting \citep{overfit} while providing no remedy against this effect. Fig.~\ref{fig:xgb-fore} presents XGBoost's predictions for \textnormal{all training-testing cases across each} class-imbalance handling \textnormal{technique}. While most scores are acceptable, results also show certain training-testing configurations leading to inflated counts of false positives, or false alarms. This \textnormal{is most} prominent when SC 24 serves as the training set (alone or in combination with another SC), and when training is done with SC 23 and tested using SC 22. Although the number of false positives \textnormal{is} high, the model still may have potential in ``all-clear'' prediction efforts, given its high recall values / low false negative values.}

\label{sec:discuss-reliable}

\begin{figure}[b]
    \centering 
        \centering \includegraphics[width=0.9\linewidth]{figures/xgb-fore.pdf}%
        \caption{\centering \textbf{Predicted TP, TN, FP, and FNs} made using an XGBoost-based model for different training and testing configurations (the x-axis shows which SC is the training set, then which SC it is tested on.) The horizontal dotted lines are the target TPs as observed per cycle (SC 22: 182, SC 23: 243, SC 24: 113).}
        \label{fig:xgb-fore}
\end{figure}

\textnormal{Additionally, we would like to discuss what methods can potentially be implemented (none tested in this paper) to avoid overfitting. \citet{c-gamma} and \citet{cv-overfit} discuss the most popular methods to reduce overfitting in SVMs: by adjusting \texttt{C} and \texttt{$\gamma$} parameters, using resampling to estimate model accuracy, and to refrain interactions with a validation data set. We apply all of these to our model through grid searches, specifying \texttt{$\gamma$ = scale}, using a k-fold CV to optimize the model by training and testing on different data subsets, and always specifying an SC as our test set. \textnormal{However,} as noted above, the SVM does not particularly seem to experience this problem. On the other hand, the overfitting we see using XGBoost may come from many different factors. Confining our grid search here to only the model's \texttt{learning rate}, and specifying gradient \textnormal{boosting via} trees, we do not augment multiple parameters associated with the algorithm. \citet{main-xgb} discuss how in terms of generalizing new data, tuning complexity \textit{and} regularization parameters may show improvements. Important parameters to consider \textnormal{tuning in XGBoost} in addition to the ones adjusted \textnormal{may be}: \texttt{max\_depth} (controls tree depth), \texttt{subsample} (specifies the number of observations to consider in each tree), \texttt{colsample\_bytree} (specifies the fraction of features to consider per tree), and setting L1 and L2 regularization terms using parameters \texttt{lambda}, and \texttt{alpha} respectively. Tuning these parameters was not evaluated in this work.}

\section{Summary and Conclusions} \label{sec:summ}
\textnormal{In this work, we have investigated the problem of predicting solar proton events (SPEs) using ML-driven algorithms, and the cross-cycle transferability of the developed models. We conclude that the XGBoost algorithm produces a finer predictive model compared to SVMs across all evaluations considered: each model with its default parameters, each model with imbalance-handling class weights applied, and each model using data oversampled by standard, SMOTE, and ADASYN separately. We also find that these class-imbalance treatment techniques lead to approximately the same results with a slight preference towards oversampling done with synthetic data generation, especially in the case of XGBoost. A summary of our key results is as follows:}
\begin{itemize}
    \item We \textnormal{built two catalogs during this effort using GOES data from SCs 22 -- 24. The first catalog records \textnormal{proton} flux features of all SPEs detected during this time. The second catalog contains daily statistical flux features of \textnormal{these high energy ($\geq$10 MeV) protons exceeding $\geq$10\, pfus, and that of soft X-rays, which serve} as the input data set for ML-based SVM and XGBoost algorithms.} 
    \item \textnormal{We use Gini Importance, Fisher scoring, and XGBoost's inherent feature ranking method to determine which flux features are most important when building our prediction models. We find the top 5 ranked features to be the same across all methods for every SC considered, though not in the same order. The last proton flux count from the previous day is the most important; as can be expected. Of all flux features retained, those of protons were more relevant than those relating to both short and long-wavelength SXRs.}
    \item \textnormal{We compare model-inherent class imbalance-handling techniques (SVM: \texttt{balanced} class weights, XGBoost: \texttt{scale positive} weights) to oversampled versions of our data set using standard duplication, SMOTE, and ADASYN oversampling. For both models, we see maximum scores obtained when ADASYN oversampling is applied (see \textnormal{Fig.} \ref{fig:over-meth})}.
    \item \textnormal{Fig.} \ref{fig:scores-use} shows TSS, ${HSS}_{2}$, \textit{and} recall scores from using XGBoost being statistically higher with respect to SVM scores for most of the considered experiments. On average, XGBoost predictions generate TSS $\sim$+0.10, ${HSS}_{2} \sim$+0.20, \textit{and} recall $\sim$+0.10 compared to those obtained by an SVM.
    \item We assess our models considering both long (using two solar cycles for training) and short (using a single solar cycle for training) timescales. We find TSS and ${HSS}_{2}$ to be comparable in both cases, for both models, for each tested solar cycle. 
    \item \textnormal{W}e compare our results to SWPC daily probabilistic forecasts and a persistence model, shown in \textnormal{Fig.} \ref{fig:pm_swpc}, and find that XGBoost (optimized with respect to TSS) outperforms these baseline models concerning TSS and recall. While the HSS$_{2}$ is higher for the persistence model (0.65 - 0.68), we note that some experiments for XGBoost without oversampling still demonstrated higher scores (0.74 - 0.75; see \textnormal{Fig.~\ref{fig:over-meth}) with comparable TSS scores (0.64 - 0.65) \textnormal{for these cases}. This indicates that ML-driven models based on operational data can outperform the current operational forecasts.
    \item Training done with SC 24 produces weaker TSS and HSS$_{2}$, even when paired with SC 22 or SC 23. Inadequate performance of single-SC training based on SC 24 may potentially indicate XGBoost's issues with overfitting, given poorer statistics of SPEs during this cycle. We also observe that TSS for SC 24 are typically lower with respect to those obtained for other cycles.
    \item Even with the highest predicted TSS scores, XGBoost produces a significant number of FPs in several training-testing configurations (Fig. \ref{fig:xgb-fore}), including all experiments involving training on SC 24. Still, the model may have potential in ``all-clear'' prediction efforts, given its high recall value.}
\end{itemize}
\textnormal{From these results, we conclude that XGBoost ensemble-based models combined with class-imbalance treatment (with no clear preference for any tested treatment) show the most potential compared to SVMs when building an SPE predictive model, in both single and double-SC trained cases. We find these ML-based models to outperform both the persistence and SWPC NOAA forecasts in all training-testing experiments considering only SCs 22 -- 24. We also claim that the expected performance of the models may differ depending on the properties of the individual solar cycle, specifically indicating that training with events during solar cycle 24 may lead to poor model performance.}\newline

\textnormal{While TSS and HSS$_{2}$ have shown slight increases across SCs when changing the aforementioned input of proton and SXR flux data, there remains a large disparity when aiming for a reliable SPE forecasting algorithm. We hypothesize further improvement in predictions by complementing our work with (1) Further hyperparameter tuning or regularization (e.g. using L1/4 Regularization Methods \citep{reg-l}, using Bayesian Regularisation \citep{reg-b}, etc.) with respect to specific metrics, (2) alternative approaches with ML algorithms such as using neural networks \citep{cnn}, logistical regression, random forests \citep{rflr}, etc. and (3) using proton and SXR fluxes' time series directly- instead of utilizing their statistical moments. We also hypothesize significant contributions to better predictive scores by considering additional input data such as properties and topologies of source ARs \citep{russel}, parameters of coronal mass ejections \citep{torres2022SpWea}, or various dynamic features of the solar corona \citep{FORWARD-intro}.}

\acknowledgements
We thank the anonymous referee for the valuable suggestions which significantly improved the quality of the manuscript. The research was supported by NASA Early Stage Innovation program grant 80NSSC20K0302, NASA LWS grant 80NSSC19K0068, NSF EarthCube grant 1639683, and NSF grant 1835958. VMS acknowledges the NSF FDSS grant 1936361 and NSF grant 1835958. EI acknowledges the RSF grant 20-72-00106.
 
\newpage
\bibliographystyle{aasjournal}
\bibliography{MAIN}{}
 
\end{document}